\documentclass[lettersize, journal]{IEEEtran}
\usepackage{xcolor}
\usepackage{colortbl}
\usepackage{threeparttable}
\usepackage{subfigure}
\usepackage{booktabs}
\usepackage{amssymb}
\usepackage{steinmetz}
\usepackage[hidelinks, urlcolor=black, colorlinks=true, linkcolor=blue, citecolor=blue]{hyperref}
\usepackage{url}
\usepackage{orcidlink}
\usepackage{cite}
\usepackage{siunitx}
\usepackage{acronym} \acrodef{3GPP}[3GPP]{3rd Generation Partnership Project}
\acrodef{AI}[AI]{Artificial Intelligence}
\acrodef{5G}[5G]{5th generation}
\acrodef{6G}[6G]{6th generation}
\acrodef{B5G}[B5G]{beyond 5G}
\acrodef{RF}[RF]{radio frequency}
\acrodef{SISO}[SISO]{single-input single-output}
\acrodef{MIMO}[MIMO]{multiple-input multiple-output}
\acrodef{CDF}[CDF]{cumulative distribution function}
\acrodef{QoS}[QoS]{quality of service}
\acrodef{ULA}[ULA]{uniform linear array}
\acrodef{ARTS}[ARTS]{Augmented ray-tracing post-processing}
\acrodef{WCG}[WCG]{wireless channel generator}
\acrodef{RAN}[RAN]{radio access network}
\acrodef{SDN}[SDN]{software-defined networking}
\acrodef{DT}[DT]{digital twin}
\acrodef{NDT}[NDT]{network digital twin}
\acrodef{RT}[RT]{ray tracing}
\acrodef{RTR}[RTR]{ray tracer}
\acrodef{CM}[CM]{coverage map}
\acrodef{Tx}[Tx]{transmitter}
\acrodef{Rx}[Rx]{receiver}
\acrodef{ISAC}[ISAC]{integrated sensing and communications}
\acrodef{NMSE}[NMSE]{normalized mean square error}
\acrodef{TRP}[TRP]{transmission reception point}
\acrodef{URA}[URA]{uniform rectangular array}
\acrodef{ID}[ID]{identification}
\acrodef{MPC}[MPC]{multipath component}
\acrodef{WB}[WB]{wideband}
\acrodef{NB}[NB]{narrowband}
\acrodef{LOS}[LOS]{line-of-sight}
\acrodef{NLOS}[NLOS]{non line-of-sight}
\acrodef{SD}[SD]{standard deviation}
\acrodef{SNR}[SNR]{signal-to-noise ratio}
\acrodef{SOH}[SOH]{spatial object handler}
\acrodef{OFDM}[OFDM]{orthogonal frequency division multiplexing}

\ifCLASSINFOpdf
\else
\usepackage[pdftex]{graphicx}
\fi

\usepackage{amsmath}
\usepackage{hhline}
\usepackage{pifont}
\usepackage{amsfonts}
\usepackage{makecell}
\usepackage{multirow}
\usepackage{algorithm, algpseudocode}

\newcommand{\Rseq}{\tilde{\mathbf{R}}}  
\newcommand{\Pseq}{\tilde{\mathbf{P}}}  
\newcommand{\Rseqnew}{\tilde{\mathbf{R}}_{\textrm{a}}}  
\newcommand{\Rseqground}{\tilde{\mathbf{R}}_{\textrm{g}}}  
\newcommand{\Rseqsimpler}{\tilde{\mathbf{R}}_{\textrm{s}}}  

\newcommand{\Sseq}{\widetilde{\mathbf{S}}}  
\newcommand{\Sseqground}{\widetilde{\mathbf{S}}_{\textrm{g}}}  
\newcommand{\Sseqnew}{\widetilde{\mathbf{S}}_{\textrm{s}}}  

\makeatletter
\renewcommand{\@cite}[2]{{\textcolor{blue}{[#1\if@tempswa, #2\fi]}}}
\makeatother

\begin{document}
%
\title{Accelerating Ray Tracing-Based Wireless Channels Generation for Real-Time Network Digital Twins}
%
%
%

\author{Cláudio~Modesto\raise0.5ex\hbox{\orcidlink{0009-0005-5184-8030}},~Lucas~Mozart\raise0.5ex\hbox{\orcidlink{0009-0000-2092-4399}}, Pedro~Batista\raise0.5ex\hbox{\orcidlink{0000-0002-0776-2054}}, André~Cavalcante\raise0.5ex\hbox{\orcidlink{0000-0003-1383-7881}}, \IEEEmembership{Member,~IEEE}, \\ and Aldebaro~Klautau\raise0.5ex\hbox{\orcidlink{0000-0001-7773-2080}}, \IEEEmembership{Senior Member,~IEEE} 
    \thanks{Cláudio Modesto, Lucas Mozart, and Aldebaro Klautau are
        with LASSE - 5G and IoT Research Group, Federal University of Pará (UFPA), Belém 66075-110, Brazil (e-mail:~\{claudio.barata,~ lucas.souza.carvalho\}@itec.ufpa.br; aldebaro@ufpa.br)}
    \thanks{Pedro Batista is with Ericsson Research, Research Area AI, Ericsson AB,~164 80 Stockholm,~Sweden (e-mail: pedro.batista@ericsson.com).}
    \thanks{André Cavalcante~is~with Ericsson Research, Ericsson Telecomunicações Ltda., Indaiatuba 13337-300, Brazil (e-mail: andre.mendes.cavalcante@ericsson.com).}
}

%
%

\markboth{PUBLISHED ON IEEE OPEN JOURNAL OF THE COMMUNICATIONS SOCIETY}{}
%



\maketitle
\begin{abstract}
\Ac{RT} simulation is a widely used approach to enable modeling wireless channels in applications such as network digital twins. However, the computational cost to execute \ac{RT} is proportional to factors such as the level of detail used in the adopted 3D scenario. This work proposes \ac{RT} pre-processing algorithms that aim at simplifying the 3D scene without distorting the channel, by reducing the scenario area and/or simplifying object shapes in the scenario. It also proposes a post-processing method that augments a set of \ac{RT} results to achieve an improved time resolution. These methods enable using \ac{RT} in applications that use a detailed and photorealistic 3D scenario while generating consistent wireless channels over time. Our simulation results with different urban scenarios scales, in terms of area and object details, demonstrate that it is possible to reduce the simulation time by more than 50\% without compromising the accuracy of the multipath \ac{RT} parameters, such as angles of arrival and departure, delay, phase, and path gain. 
\end{abstract}

\begin{IEEEkeywords}
3D scenario simplification, channel augmentation, \acf{RT},  synthetic data generation.
\end{IEEEkeywords}

%
\IEEEpeerreviewmaketitle

\section{Introduction}
\IEEEPARstart{O}{ver} the past few years, \acf{RT} simulations have gained prominence in wireless communications, complementing traditional stochastic and geometry-based stochastic channel models~\cite{almers2007, Danping2019}. \ac{RT} has emerged as a key enabler for various research domains, driving advancements in THz communications~\cite{Jianhua2024, Yi2024}, \ac{ISAC}~\cite{Arnold2022}, and \acp{NDT}~\cite{Yiwen2021, Ruah2024}.  
Recognizing its potential, standardization bodies are actively integrating \ac{RT} into future network frameworks. The \ac{3GPP}, for instance, is exploring its role in \ac{ISAC} development for 6G~\cite{R12400691}, while the O-RAN Alliance~\cite{oran_digital_twin_2025} investigates its application in network planning through digital twins.

The main reason for the ubiquitous presence of this technology is the ability to generate high-fidelity channel parameters~\cite{Klautau2018}, such as \acp{MPC} and channel impulse response,  using a \emph{site-specific} approach. However, the computational cost of achieving this level of fidelity is the main drawback associated with \ac{RT}. 
This cost depends on 
the detail level of the 3D scenarios~\cite{Novak2021, Mozart2024} and several other aspects, such as
number of \acp{TRP}, maximum depth of interactions, and types of supported propagation phenomena (reflection, diffraction, etc.)~\cite{Zentner2013, zhu2024, Hussain2017}.

This article presents methods to speed up the estimation of a temporal sequence of wireless channels using \ac{RT} for applications such as real-time network digital twins~\cite{Borges2024, Testolina2024, yu2025}. 
Instead of proposing a new \ac{RT} execution approach, the proposed solutions are preprocessing and post-processing algorithms that rely on a conventional ray tracer to calculate a set of reference channels. The preprocessing method helps the conventional \ac{RT} by simplifying the 3D scenario, while the post-processing method improves the time resolution of the reference channels by a given upsampling factor. In summary, the main contributions of this article are:
\begin{itemize}
 
\item We pose two problems that help directing research on speeding up \ac{RT} by describing them in a multi-objective framework.

\item A novel \ac{RT} post-processing method that augments a reference sequence of wireless channels, by considering the inherently stochastic processes of rays ``birth'' and ``death'', as well as signal processing to interpolate consecutive rays and generate a new sequence of channels with improved time resolution.

\item A new category of preprocessing methods to simplify the 3D scene used as an input to a conventional ray tracer, which alleviates the computational burden by choosing a specific area of interest, which avoids loading the ray tracer with unnecessary details.

\item Relevant experimental results obtained with a systematic and thorough methodology, which is based on the creation of a ``ground-truth'' that enables evaluating solutions according to quantitative figures of merit.
\end{itemize}

The remainder of this article is organized as follows.
Section \ref{sec:problem_formulation} describes the fundamental properties of the problems addressed, considering all the definitions and mathematical elements regarding a \ac{RT} simulation and the channel model adopted. With the required nomenclature and problems defined, Section \ref{sec: related_work} discusses the previously published related work, which helps localizing the proposed solutions with respect to the state-of-the-art. Section~\ref{sec:speeduprt} positions the two problems in the context of network digital twins, illustrating how their solutions can be integrated.
Section~\ref{sec:proposed_solution} presents the solutions to each of the posed problems, 
detailing the preprocessing and post-processing methods.
Section \ref{sec: evaluation} separately validates the two proposed solutions  in different scenarios with respect to channel characteristics and simulation time. In Section \ref{sec: conclusion} we conclude this article and point out future directions.

\section{Problems Formulation}
\label{sec:problem_formulation}

This article focuses on two distinct problems related to speeding up the generation of  a \emph{sequence} of wireless channels represented by \acp{MPC} (or rays), as the ones obtained by conventional \ac{RT} algorithms.
The first problem consists of using post-processing to obtain an upsampled version of the \ac{RT} results with improved time resolution. 
The goal is to achieve these results in significantly less time than running a conventional ray tracer software with improved time resolution, while still obtaining accurate wireless channels.

The second problem involves implementing a preprocessing stage to simplify 3D scenarios by reducing the polygon count while preserving the integrity of the respective sequence of wireless channels. 
For concreteness, it is assumed here that each 3D object is represented by a single mesh.\footnote{For instance, organized a file of type \texttt{.ply} (polygon file format).} Other representations can be used, but in this article, a 3D scene with $O$ objects is represented by a set $\cal S$ with $O$ meshes.
The primary goal of these solutions is to accelerate \ac{RT} simulations. Besides, such preprocessing methods may be crucial for enabling RT in cases where the complexity of the 3D scenario exceeds the capabilities of the RT software, and a simpler 3D scenario becomes mandatory.

Regarding the \ac{RT} software, we are interested in this article on ray tracers that model the propagation of electromagnetic waves in communication bands, which have very distinct properties when compared to ray tracers that are used to model light propagation in 3D rendering software, such as in~\cite{woong2022, parker2010}.
And without loss of generality, we will assume one \ac{Tx} / \ac{Rx} pair. But the solutions discussed in this article can be used 
for scenarios with multiple \acp{TRP}.

The following block diagram summarizes the assumed conventional \ac{RT} pipeline:
\begin{center}
   $\mathcal{S}$ $\rightarrow$ \fbox{\mbox{ray tracer}} $\rightarrow$ $\mathcal{R}$ 
    $\rightarrow$ \fbox{\mbox{signal processing}} $\rightarrow$ $\mathbf{H}$ 
\end{center}
where a ray tracer calculates a set of rays $\mathcal{R}$ 
representing the   
wireless channel, given the 3D scene described by meshes in $\mathcal{S}$.
It is then assumed that $\mathcal{R}$ provides all information needed to obtain the desired representation of the wireless channel. For instance, the experiments in this article assume that $\mathbf{H}$ could assume a \ac{NB} or \ac{WB} \ac{MIMO} \emph{channel matrix}~\cite{Heath2016}, obtained via the \emph{geometric model} applied to the results of a \ac{SISO} \ac{RT} simulation~\cite{Trindade2018}.

It should be noted that the two posed problems can be considered \emph{supervised}, in the sense that it is assumed there is a ``ground-truth'' to be used in quantitative (not only qualitative) comparisons. This ground-truth is obtained by invoking the conventional ray tracer for each scene ${\cal S}_m^\textrm{g}$ in the sequence $\Sseqground = [{\cal S}_m^\textrm{g}]_{m=1}^M$ with the highest time resolution (that corresponds to a sampling interval $T_{\textrm{s}}$). This ground truth must be properly interpreted, especially because distinct ray tracers may lead to very different results~\cite{zhu2024}. It is out of the scope of this article to assess how accurate the \ac{RT} results are with respect to measurements~\cite{Gotszald2015}. 
The following subsections describe the two problems.

\subsection{Post-processing RT for improved time resolution}
Assume the simulation of a communication scenario in which a site-specific channel is obtained via \ac{RT} for each 3D \emph{scene}~\cite{Klautau2018}. 
A 3D scene is composed of fixed and mobile 3D objects. Mobile objects are \ac{Tx}, \ac{Rx} and scatterers (for instance, vehicles on a street). The dynamics along the sequence of scenes over discrete-time $n=1, \ldots, N$ are captured through periodic sampling with a sampling interval $T_{\textrm{o}}$, such that the $n$-th scene corresponds to time $t=(n-1) T_{\textrm{o}}$ seconds.\footnote{Public 
wireless channel datasets created with this methodology can be found at \url{https://www.lasse.ufpa.br/raymobtime}.}
Invoking the ray tracer with this relatively ``low'' time resolution leads to the sequence $\Rseq=[ \mathcal{R}_1, \mathcal{R}_2, \ldots, \mathcal{R}_n, \dots, \mathcal{R}_N ]$, where $\mathcal{R}_n$ corresponds to a set of rays representing the   
wireless channel at discrete time $n$, as obtained from scene ${\cal S}_n$. 
Without loss of generality, this description assumed that $N$ \ac{RT} runs were necessary to obtain 
$\Rseq$, but the ray tracer may support special mobility features and obtain $\Rseq$ with less than $N$ runs.

\begin{figure}[htp]
    \centering
    \includegraphics[scale=0.38]{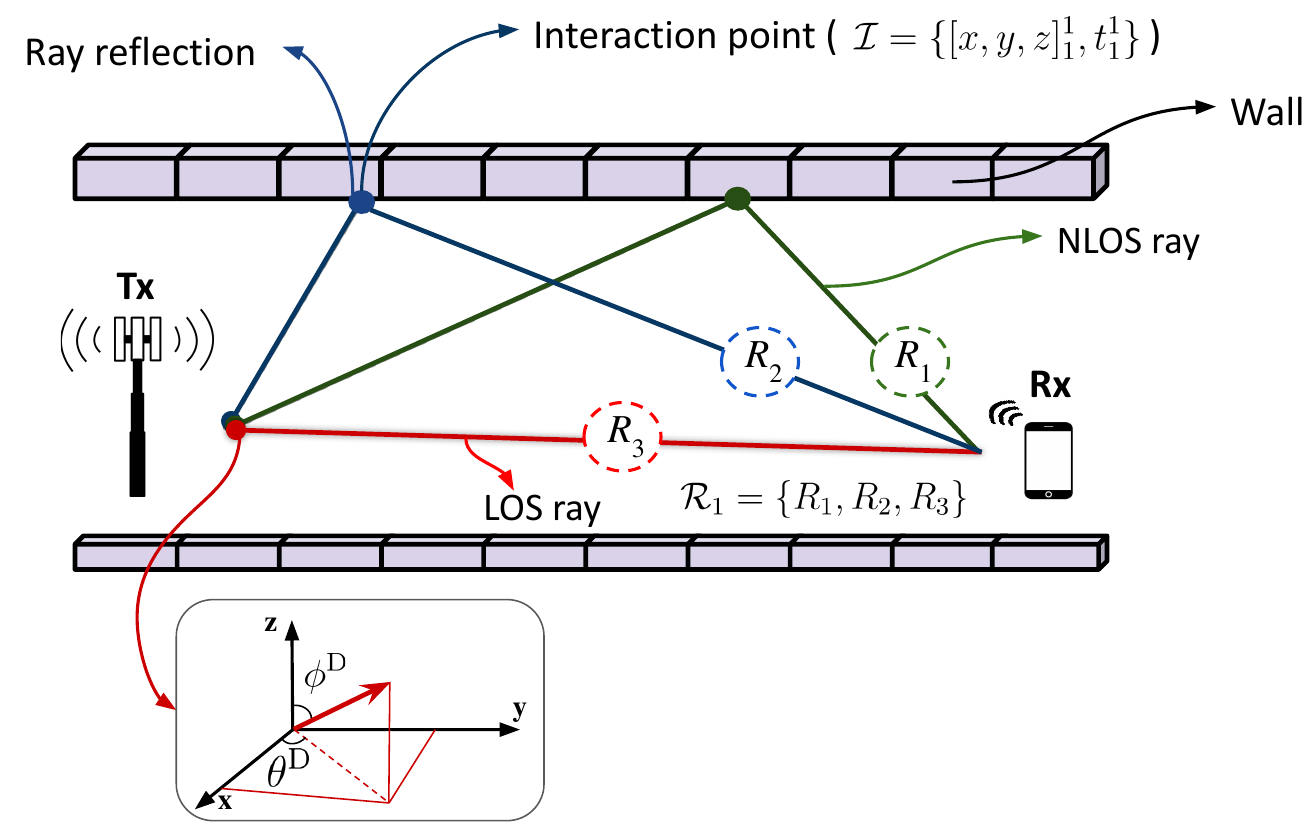}
    \caption{An illustrative description of the adopted nomenclature. The “blocks” represent brick walls that impose interactions across the ray trajectories. In this case, we have at discrete time $n = 1$, a set of rays $\mathcal{R}_1$ with $K_1 = K_\textrm{max} = 3$. The ray $R_3$ represents a \ac{LOS} ray; $R_1$ and $R_2$ are \ac{NLOS}. For each ray $R$, the 3-tuple with $\mathbf{p}$, $b$, and $\mathcal{I}$ is calculated by the ray tracer.} 
    \label{fig:rt_wo_mobility_definitions}
\end{figure}

It is assumed that, for the run corresponding to the $n$-th 3D scene, the adopted ray tracer returns the information respective to a single channel organized as a set of $K_n$ rays $\mathcal{R}_n = \{R_1^n, R_2^n, \ldots, R_r^n, \ldots, R_{K_n}^n\}$, where $R_r^n$  denotes the $r$-th ray, as indicated in Fig.~\ref{fig:rt_wo_mobility_definitions}, which suppresses the superscript identifying scene $n$. 
The number of rays $K_n \le K_{\textrm{max}}$ can vary across scenes and is upper-bounded by a maximum value $K_{\textrm{max}}$ specified to the ray tracer.
To simplify notation, when $n$ and $r$ can be made implicit, we denote a ray $R_r^n$ as $R$.

Each ray $R=(\mathbf{p}^n_r, b^n_r, \mathcal{I}^n_r)$ is a tuple of three elements: a vector $\mathbf{p}^n_r \in \mathbb{R}^7$ with the numerical parameters listed in Table \ref{tab:ray_parameters}; a binary variable $b^n_r$ indicating whether the ray is \ac{LOS} or \ac{NLOS}, and a set $\mathcal{I}^n_r$ describing all \emph{interactions} between the ray and the objects composing the 3D scene.
And each element of the set $\mathcal{I}^n_r$ is a tuple with two elements: a vector $\mathbf{x} = (x,y,z)$ informing the 
position in meters of the interaction, with respect to the scene origin $(0,0,0)$; and an integer $t$ indicating the type 
of propagation phenomenon (reflection, diffraction, diffusion scattering, refraction / 
transmission, etc.) undergone by the ray at the given interaction.

\begin{table}[htp]
    \centering
    \caption{Parameters in vector $\mathbf{p}$ describing a single ray (\ac{MPC})}
    \scalebox{1.1}{\begin{tabular}{cc}
    \toprule
    Parameter & Description \\
    \midrule
    $\alpha$ &  Gain (real-valued magnitude) \\
    $\Omega$ & Phase (radians) \\
    $\tau$ & Delay (seconds) \\
    $\theta^\textrm{D}$ & Azimuth of departure (AoD) angle (radians) \\
    $\phi^\textrm{D}$ & Elevation of  departure (EoD) angle (radians) \\
    $\theta^\textrm{A}$ & Azimuth of arrival (AoA) angle (radians) \\
    $\phi^\textrm{A}$ & Elevation of arrival (EoA) angle (radians) \\
    \bottomrule
    \end{tabular}}
    \label{tab:ray_parameters}
\end{table}

Hence, all $I_r^n$
interactions of the ray $R_r^n$ are denoted (see Fig.~\ref{fig:rt_wo_mobility_definitions}) as 
$\mathcal{I}_r^n = \{ (\mathbf{x}_r^n[1], t^n_r[1]), \ldots,  (\mathbf{x}_r^n[I_r^n], t^n_r[I_r^n]) \}$.
It is assumed that a channel can have at most one \ac{LOS} ray ($b=1$) and for this specific ray, the interactions set $\cal I$
is empty.
We refer to a ray tracer that fully encapsulates all the information within
$\mathcal{R}_n$, including the parameters in Table~\ref{tab:ray_parameters} and the interaction set $I_r^n$, as \emph{conventional}.
Examples of conventional ray tracers are Remcom's Wireless InSite\textsuperscript{\text\textregistered}~\cite{wirelessInsite} and NVIDIA's  Sionna\textsuperscript{\texttrademark}~\cite{Hoydis2023}.

It is also assumed a distance measure $D(\mathbf{H}, \mathbf{\hat H})$ that can be used as a figure of merit when evaluating the accuracy of representing the channel $\mathbf{H}$ by an estimate $\mathbf{\hat H}$. The distance $D(\mathbf{H}, \mathbf{\hat H})$ adopted in this article is the \ac{NMSE}, which is defined in Section~\ref{sec: evaluation}.
With a slight abuse of notation, and recalling that each channel $\mathbf{H}$ is deterministically derived from its corresponding set of rays $\mathcal{R}$, an \emph{average distance} between two sequences $\Rseq = [ \mathcal{R}_m ]_{m=1}^M$ and $\Pseq = [ \mathcal{P}_m ]_{m=1}^M$ of sets of rays is defined as
\begin{equation}
\overline{D}(\Rseq, \Pseq) = \frac{1}{M} \sum_{m=1}^M D(\mathbf{H}_m^{\textrm{R}}, \mathbf{H}_m^{\textrm{P}}),    
\end{equation}
where $\mathbf{H}_m^{\textrm{R}}$ and $\mathbf{H}_m^{\textrm{P}}$ are the corresponding channel matrices for sets of rays $\mathcal{R}_m$ and $\mathcal{P}_m$, respectively.

Assuming an integer $U > 1$ \emph{upsampling} factor, the augmentation problem is to create $U-1$ sets of rays located in time between consecutive pairs of the input sequence $\Rseq$ obtained with a conventional ray tracer. Hence, the complete augmentation procedure creates $(N-1)(U-1)$ new 
sets of rays that, when properly merged with the original ray sets in $\Rseq$, compose the new augmented sequence $\Rseqnew$ with a total of $M=N \times U - (U-1)$ sets of rays.\footnote{The total number of sets in $\Rseqnew$ would be $M=N \times U$ in case $t \in [0, N \times T_{\textrm{o}}]$~\si{s} but we want to avoid having to create $U-1$ new channels after the last channel in $\Rseq$.}

The ground truth for this problem is obtained from the sequence $\Sseqground$ of $M$ scenes, in which consecutive scenes are separated in time by the target sampling interval $T_{\textrm{s}}$. Invoking the conventional ray tracer on this sequence of scenes creates the ground truth sequence $\Rseqground$ such that $\Rseq$ corresponds to a decimated version of $\Rseqground$ by a factor of $U$. Obtaining $\Rseqground$ can take considerable time because $T_{\textrm{s}}$ is relatively small and each scene in $\Sseqground$ may have a large count of polygons.

\subsection{Preprocessing RT 3D input}

\begin{figure}[htp]
    \centering
    \includegraphics[scale=0.17]{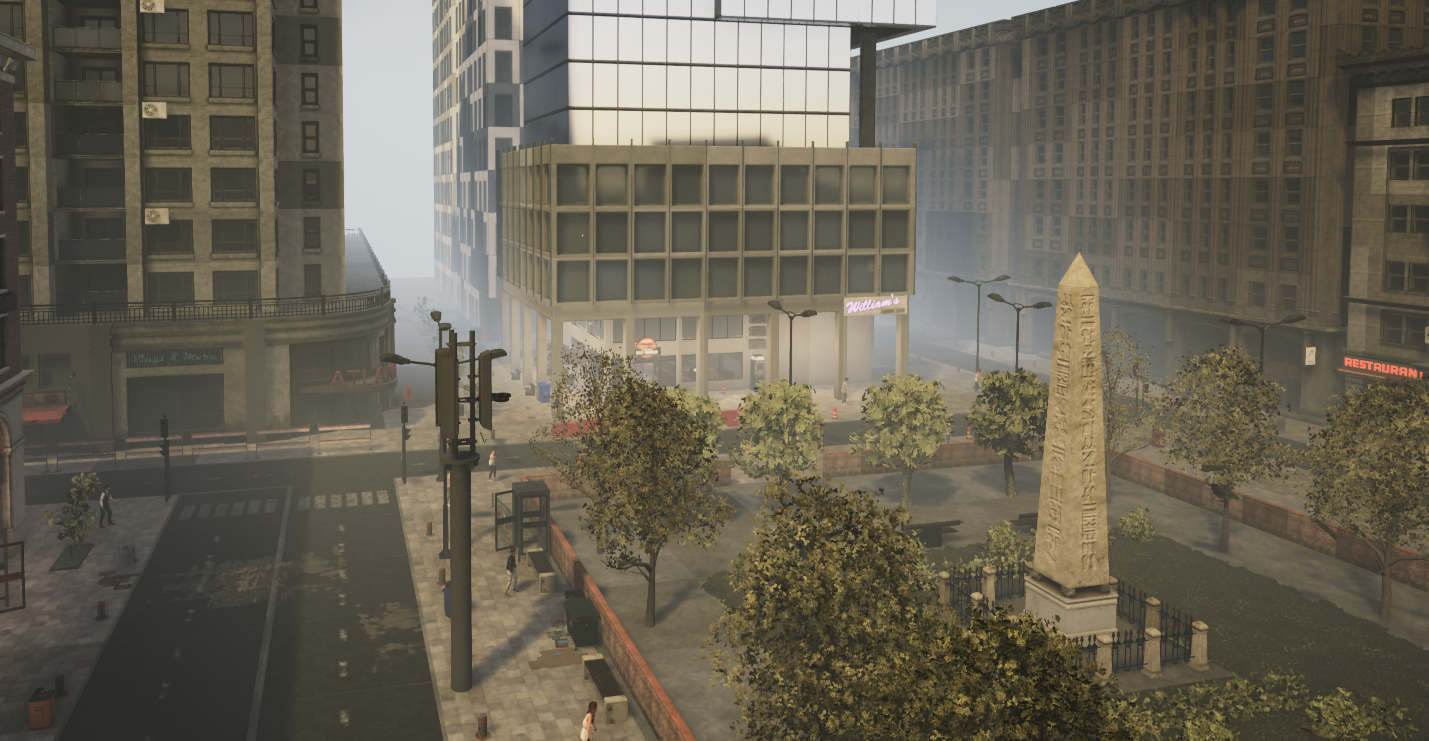}
    \caption{Scenario called Modern City rendered with Unreal engine.}
    \label{fig:original_scenario}
\end{figure}

Another problem investigated in this article is how to enable or speed up \ac{RT} by simplifying the 3D scenes without compromising accuracy and important features of channel sequences such as \emph{spatial consistency}~\cite{3gppTR38901}. Engines such as Unreal~\cite{unrealengine} and Unity~\cite{unityUnityRealTime} support scenes with a number of faces that are orders of magnitude larger than the number supported by existing ray tracer software, as depicted in Fig.~\ref{fig:original_scenario}. Hence, in spite of solutions to P1 using \ac{RT} augmentation, executing \ac{RT} itself can be unfeasible for extremely detailed 3D scenarios.

An important figure of merit when solving this problem is the computational cost $C(\mathcal{R})$ of obtaining the set of rays $\mathcal{R}$ that corresponds to channel $\mathbf{H}$ by executing the conventional ray tracer from the input scene described by $\mathcal{S}$. This cost can be defined as a weighted average among several performance indicators, such as memory consumption $c_1$, processing time $c_2$, and data transfer latency $c_3$. For instance, the cost 
could be $C(\mathcal{R})=2 c_1 + 0.3 c_2 + c_3$. Defining $C(\mathcal{R})$ depends on the application and is not considered in this work,
but it is important that it measures the total time to obtain each $\mathcal{R}$. The complete pipeline to obtain each $\mathcal{R}_m$ varies, and can be composed of a preprocessing step, augmentation (post-processing) and/or a direct conventional \ac{RT}.
Assuming the defined $C(\mathcal{R})$ represents the total cost for obtaining $\mathcal{R}$, one can calculate the \emph{average cost}
\begin{equation}
\overline{C}(\Rseq) = \frac{1}{M} \sum_{m=1}^M C(\mathcal{R}_m)    
\end{equation}
for computing a sequence $\Rseq = [ \mathcal{R}_m ]_{m=1}^M$ of $M$ sets of rays.

The solution to P2 should correspond to a sequence of scenes $\Sseqnew$ that is simpler than the original sequence $\Sseq$, in the sense 
that $\overline{C}(\Rseqsimpler) < \overline{C}(\Rseq)$, where $\Rseqsimpler$ and $\Rseq$ are the sequences of rays sets corresponding to $\Sseqnew$ and $\Sseq$, respectively. The accuracy of a P2 solution can be measured by the average distance $\overline{D}(\Rseq, \Rseqsimpler)$.

Some solutions may use information not only from scene $\cal S$ to simplify it, but also from other (neighbor) scenes and even \ac{RT} results of some specific scenes. If \emph{causal} processing is required, when simplifying the $m$-th scene, an \emph{inter-scenes} solution to P2 uses information from previous scenes ${\cal S}_n$ at time $n < m$ and/or previously calculated \ac{RT} results. An \emph{intra-scene} solution is the special case in which, at time $m$, the P2 solution uses only information
from the sets of meshes described in current scene ${\cal S}_m$.

\subsection{Multi-objective framework to define the two problems}

In \ac{RT} applications such as network digital twins, there exists a trade-off between accuracy and computational cost. The figures of merit 
$\overline{D}(\Rseq, \Pseq)$ and $\overline{C}(\Rseq)$ were introduced to help define the two problems in a multi-objective framework. 
The overall goal is to identify solutions on the \emph{Pareto front}, which optimally balance these competing objectives~\cite{sharma2022}. 
Based on the mentioned concepts, the first problem is defined as follows.

\textbf{P1 (\ac{RT} Augmentation)}:  
Given a sequence of ray sets \( \Rseq \) and \emph{upsampling} factor $U>1$, determine an efficient method, as measured by the average cost $\overline{C}(\Rseqnew)$, to generate an augmented sequence \( \Rseqnew \) that represents wireless channels over the time range \( t \in [0, (N-1)T_{\textrm{o}}] \)~\si{s}, with a sampling interval \( T_{\textrm{s}} = T_{\textrm{o}}/U \), with the augmented sequence \( \Rseqnew \) maintaining high fidelity to the ground truth channels \( \Rseqground \), as measured by the average distance \( \overline{D}(\Rseqnew, \Rseqground) \).  
\hfill $\square$

Considering the more general non-causal inter-scenes case, the second problem described in this article is defined as follows.

\textbf{P2 (3D scenario simplification)}:
Given a sequence of scenes $\Sseq$ and eventually the \ac{RT} results from specific scenes, obtain a new sequence
$\Sseqnew$ such that the average cost $\overline{C}(\Rseqsimpler)$ of running the conventional ray tracer
and the average distance $\overline{D}(\Rseq, \Rseqsimpler)$ are minimized, where $\Rseqsimpler$ and $\Rseq$ are the sequences of rays sets corresponding to $\Sseqnew$ and $\Sseq$, respectively. 
\hfill $\square$

Section~\ref{sec:proposed_solution} describes solutions to problems P1 and P2. Before that, the next section discusses the related works regarding the P1 and P2 problems.

\section{Related Work}
\label{sec: related_work}
\label{sec:Related Work}

\newcommand{\xmark}{\ding{55}}%
\newcommand{\cmark}{\ding{51}}%

\begin{table*}[htbp]
\caption{Works that are related to problems P1 and P2, as well as some that improve \ac{RT} itself, which is out of the scope of this article}
\label{tab: rt_simplification_methods}
\centering
\scalebox{1}{\begin{tabular}{cccccc}
\toprule
  
\textbf{Work} & \textbf{Tested ray tracer} & \textbf{Supported ray tracer}  & \makecell{\textbf{P1}\\\textbf{\ac{RT} augmentation}} & \makecell{\textbf{P2}\\\textbf{3D scenario simplif.}} & \textbf{\makecell{Improved\\\ac{RT}}}  \\
\midrule
\makecell{Zentner and Mucalo, 2013~\cite{Zentner2013}} & Authors-developed & \textit{Conventional} ray tracers  & \cmark  & \xmark & \xmark \\ 
\midrule
\makecell{Bedford et al., 2020~\cite{Bedford2020}} & \textit{Wireless InSite} & \textit{Conventional} ray tracers & \xmark & \cmark & \xmark\\
\midrule
\makecell{Mi et al., 2020 
\cite{Mi2020}} & Unspecified & \textit{Conventional} ray tracers & \xmark  & \cmark & \xmark\\
\midrule
\makecell{Novak et al., 2021~\cite{Novak2021}} & \textit{Signal3D}~\cite{Novak2016} & \textit{Conventional} ray tracers & \xmark & \cmark & \xmark\\ 
\midrule
\makecell{Mozart et al., 2024~\cite{Mozart2024}} & \textit{Sionna} & \emph{Conventional} ray tracers & \xmark & \cmark & \xmark\\
\midrule
\makecell{Testolina et al., 2024~\cite{Testolina2024}} & \textit{Sionna} & \textit{Sionna}  & \xmark & \cmark & \xmark\\
\midrule
Our work & \textit{Sionna} & \emph{Conventional} ray tracers & \cmark & \cmark & \xmark\\
\midrule[1.5pt]
\makecell{Gotszald, 2015, \cite{Gotszald2015}} & Author-developed & -- & \xmark & \xmark & \cmark \\
\midrule
\makecell{Liu et al., 2024, \cite{Liu2024}} & Authors-developed  & -- & \xmark & \xmark & \cmark \\
\midrule
\makecell{Lecci et al., 2020 \cite{Lecci2020}} & \textit{Q-D Realization} & -- & \xmark & \xmark & \cmark \\
\midrule
\makecell{Lecci et al., 2021 ~\cite{Lecci2021}} & \textit{Q-D Realization} & --  & \xmark & \xmark & \cmark\\ 
\bottomrule
\end{tabular}}
\end{table*}

The cost to perform \ac{RT} simulation depends on several factors, but quickly increases mainly due to two conditions~\cite{Lecci2020}. The first is when the number of \acp{TRP} becomes larger, consequently growing the number of channels to be simulated per scene. And the second  is when the 3D scenario is graphically complex, in the sense of the number of polygons and meshes. To overcome these drawbacks, several authors proposed simplification methods that, in spite of not being strict solutions to P1 or P2, are useful to position the current article with respect to the state-of-the-art. The most important related methods are summarized in Table~\ref{tab: rt_simplification_methods}.

Table~\ref{tab: rt_simplification_methods} also describes the ray tracers that are supported by the proposed method, as well as the ray tracer actually used in the experiments presented in the respective paper. The table identifies that a solution to a problem closely related to P1 was proposed in~\cite{Zentner2013}. With respect to P2, five other works are listed. The table is expanded with the last rows and last column to explicitly indicate works that are outside the scope of this article. The column ``Improved \ac{RT}'' indicates a few publications about the extensively investigated problem of finding good algorithms for \ac{RT} applied to telecommunications. While this article deals with different categories of methods (\ac{RT} preprocessing and post-processing methods), for completeness, the last four papers in Table~\ref{tab: rt_simplification_methods} provide to the interested reader a small but diverse sample of \ac{RT}  algorithms.

Using a ray tracer called \textit{Q-D realization} \cite{qdrealization}, Lecci et al. \cite{Lecci2020, Lecci2021} proposed a method to speed up \ac{RT}, based on thresholding the gain and number of reflections of each ray. In \cite{Gotszald2015}, the author discusses his ray tracer and compares it to Wireless InSite, which is one of the most accurate \ac{RT} software for telecommunications~\cite{zhu2024}. The work in \cite{Liu2024} illustrates how \ac{RT} can be customized to outdoor scenarios using sub-6~GHz carrier frequencies. 
As indicated, the last four works in Table~\ref{tab: rt_simplification_methods} 
do not compete with the solutions proposed in this article, which complement them.

The work closest to ours is the one by
Zentner and Mucalo~\cite{Zentner2013}, which formulated an application that uses an interpolation method in the ray parameter domain, leveraging the idea of \emph{ray entities} that groups a set of similar
rays that undergo the same propagation phenomenon across the interactions with the objects in the scenario. 
Organizing similar rays helped 
interpolation methods such as cubic splines, linear, or polynomial, to augment the sequence originally generated
by the conventional ray tracer as suggested in problem P1.
Our solution to P1 has the following improvements with respect to \cite{Zentner2013}: a) we consider all parameters that compose a set of rays $\cal R$ while the authors in \cite{Zentner2013} did not consider the ray phase (which is the hardest parameter to estimate), but only the time of arrival, angles of departure and elevation, and received power; and b) we consider the existence of a ray birth-and-death process that was not taken into account in \cite{Zentner2013}.

Related to problem P2 of geometrical simplification, the authors in~\cite{Bedford2020, Novak2021} proposed an approach to accelerating the \ac{RT} simulation specifically in a natural cave environment (for mining scenarios).
These authors used different simplification methods, with the most relevant to this article being the so-called \emph{quadric edge collapse}, which alleviates the computational burden caused by a large number of faces.
In our previous work~\cite{Mozart2024} we extended~\cite{Novak2021} to large urban scenarios (instead of natural caves), using the built-in mesh simplification modules from Blender~\cite{blender} and considering different scales of simplification. Testolina et al.~\cite{Testolina2024} presented a framework called \textit{BostonTwin} intended to be used for \ac{RT} experimentation 
considering information about the whole city of Boston.
An important tool made available in this framework was a 3D scenario simplification method called the \emph{vertex clustering algorithm}~\cite{Low1997}, which is used as a baseline in Section~\ref{sec: evaluation}.

The mentioned solutions to P2 are based on mesh simplification. Alternative approaches were adopted
by Mi et al.~\cite{Mi2020}, which used two methods called \textit{uniform grid space partitioning} and \emph{$k$-D tree space}. Both methods share some similarities to the \emph{cut-out} based methods proposed in this article.
The former method divides the environment into grids and discards the ones (and the objects that belong to this grid) that do not interact with the \ac{RT} paths, reducing the number of ``intersection tests''. The latter method also uses the idea of grid division, but in this case, considers a non-uniform partitioning based on $k$-D tree algorithm~\cite{Berg2008}. A main drawback related to these methods is the need to divide the 3D scenario to check which grid is not contributing to the \ac{RT} simulation. In large scenarios, this process can consume a considerable amount of time.

\section{Speeding up RT in network digital twins}
\label{sec:speeduprt}

It is useful to visualize the solutions to problems P1 and P2 within a network digital twin framework centered on \ac{RT}. This allows for a better understanding of how the mentioned solutions are implemented in practice. It is assumed that the framework contains three main modules: 
the \ac{WCG} implements a solution to P1, the \ac{SOH} solves P2, and an \emph{orchestrator} organizes the optimization pipeline operation, adjusting the involved hyperparameters to achieve the target real-time factor, trading off speed and accuracy.

\begin{figure}[htpb]
    \centering
    \includegraphics[scale=0.34]{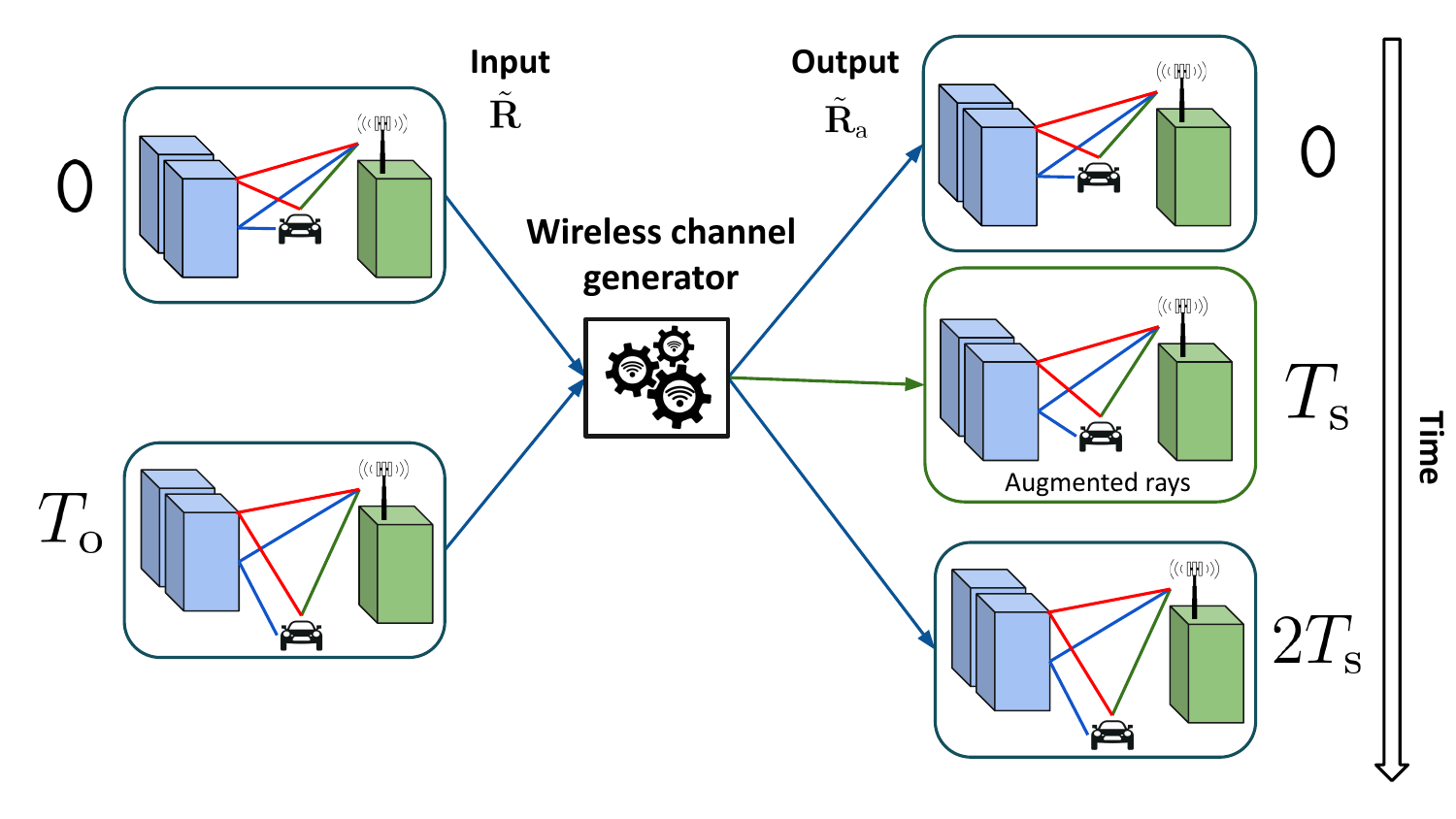}
    \caption{\ac{WCG} module working with upsampling $U=2$ and augmenting the input sequence $\Rseq$ to create $\Rseq_\text{a}$ with 
    three sets of rays.}
    \label{fig:io_wcg}
\end{figure}

The role of the \ac{WCG} module as a post-processing method (executed after \ac{RT}) is depicted in Fig.~\ref{fig:io_wcg}. In this example, \ac{WCG} augments $\Rseq$, which contains $N=2$ sets of rays, using $U=2$ to create $\mathcal{P}$ with $N \times U-(U-1)=3$ sets of rays and a corresponding output sampling interval $T_{\textrm{s}}=T_{\textrm{o}}/2$~\si{s}.
Fig.~\ref{fig:io_wcg} does not show it explicitly, but besides $\Rseq$, the \ac{WCG} module can leverage other information obtained via the orchestrator, such as descriptors of the 3D scenes.

\begin{figure}[htbp]
    \centering
    \includegraphics[scale=0.34]{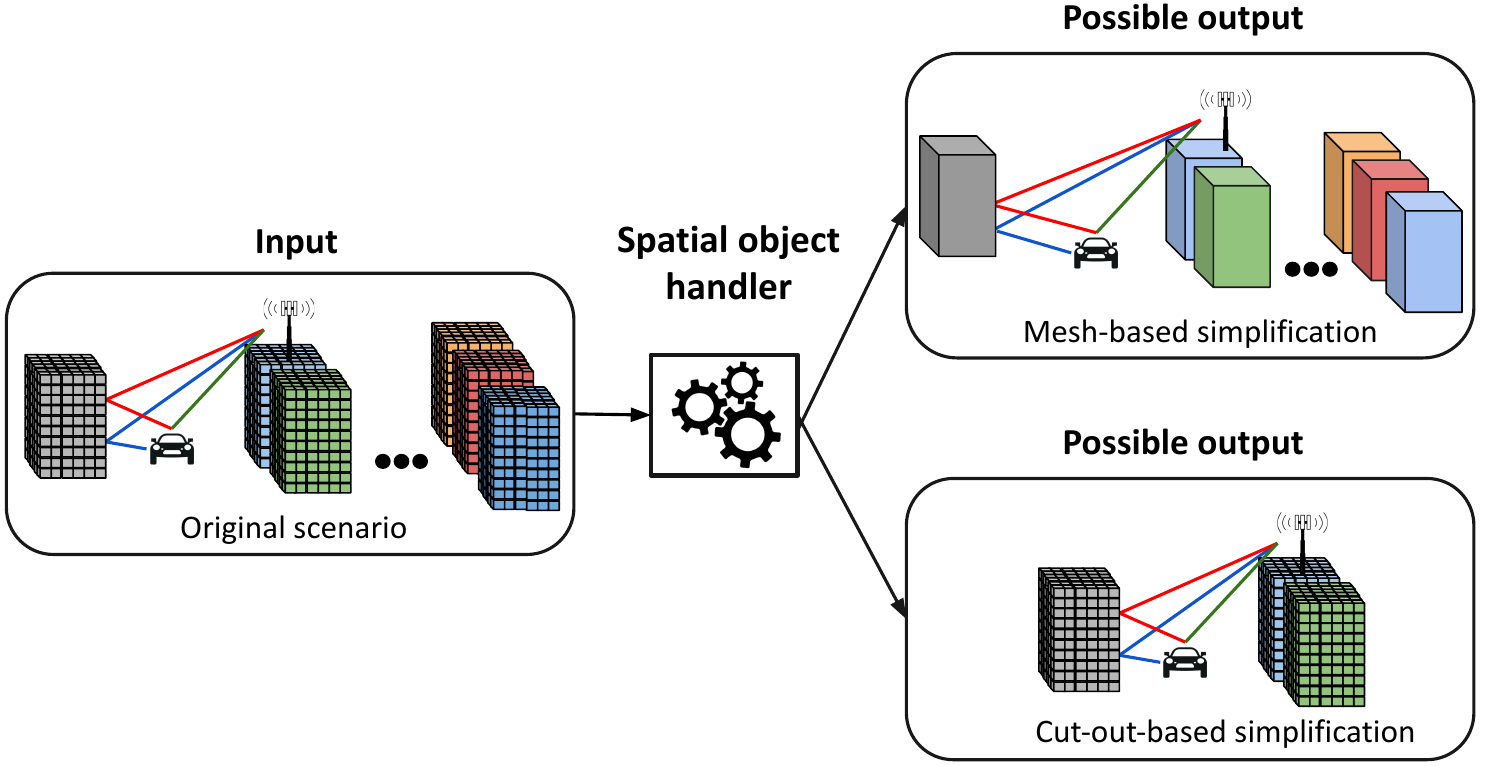}
    \caption{Input-output relation of the \ac{SOH} module. In this example, the model received a set of meshes $\mathcal{S}$ and offer two ways to simplify it, the mesh-based intended to simplify extremely detailed meshes and cut-out-based approach aimed to remove unnecessary scenario regions.}
    \label{fig:io_soh}
\end{figure}

As depicted in Fig. \ref{fig:io_soh}, the \ac{SOH} module aims at adjusting the complexity of the 3D scenario as a preprocessing method (executed before \ac{RT}). 
It helps our discussion to organize solutions to P2 into two categories:
\emph{mesh-based} and \emph{cut-out-based simplification}. The former category encompasses several classic methods that focus on reducing the mesh count and were applied in distinct applications~\cite{Cignoni1998, Berg2008}, including network digital twins~\cite{Testolina2024}. 
In contrast, the methods in the latter category focus on reducing the volume represented by the 3D scene, discarding objects that do not impact the associated \ac{RT} result. 

\begin{figure}[htb]
    \centering
    \includegraphics[scale=0.34]{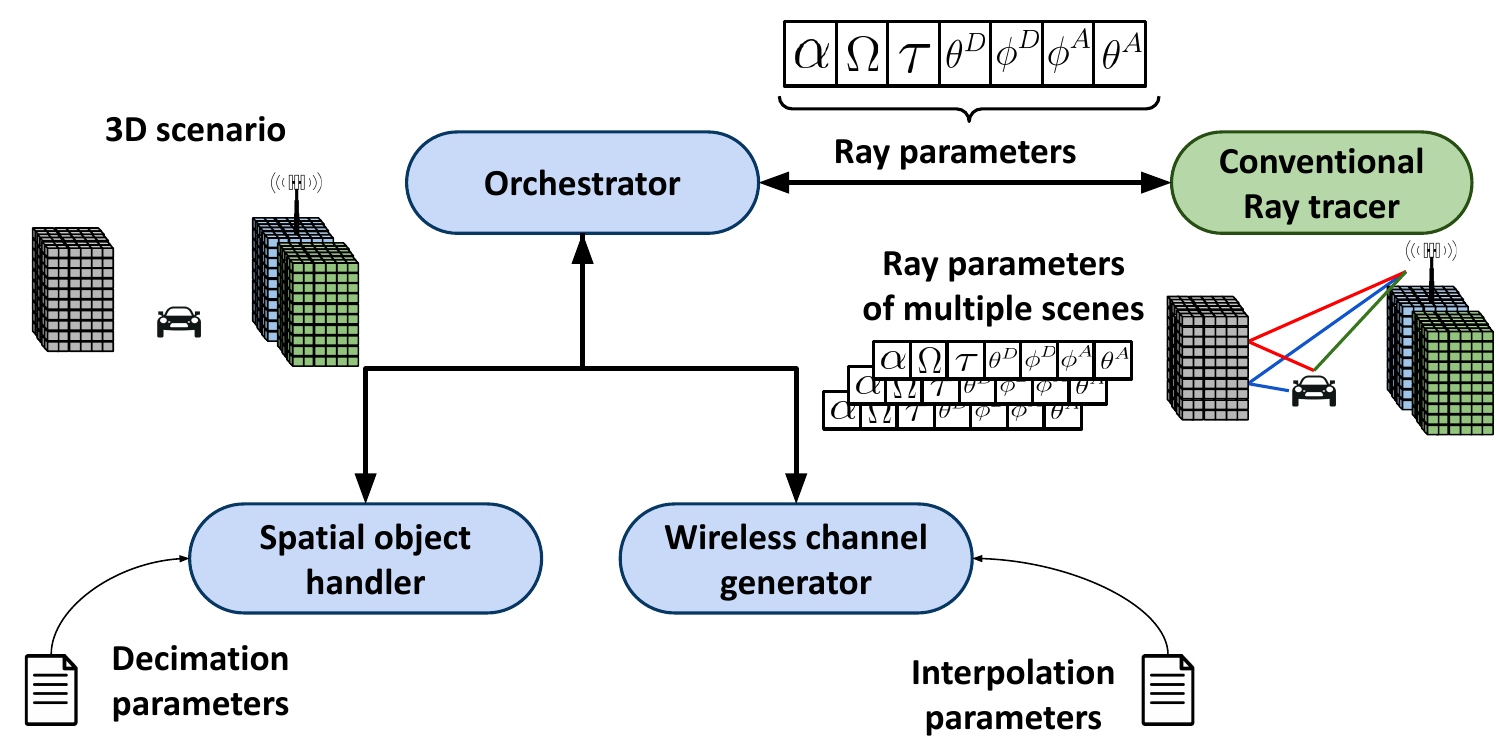}
    \caption{Relationship between all elements of our proposed framework.}
    \label{fig:proposed_framework}
\end{figure}

The methods in the category \emph{cut-out-based simplification} are novel and described in Section~\ref{sec:proposed_solution}. The main idea is that, as the simulation evolves over time, the \ac{TRP} and scatterers eventually move to new positions, and the \ac{SOH} module can reduce the effective volume (or area) to be simulated based on previous \ac{RT} runs and new object positions. The \ac{SOH} can then create and send to the ray tracer a simpler 3D scene.

The orchestrator sets the \ac{SOH} module to use a given 3D scene simplification method, and then invokes a conventional ray tracer, as depicted in Fig. \ref{fig:proposed_framework}. After \ac{RT} is concluded, the orchestrator calls the \ac{WCG} module, providing it with information composed of previous \ac{RT} results and the 3D scenario. The \ac{WCG} then generates sets of rays representing wireless channels with high time resolution.

\section{Proposed Solutions}\label{sec:proposed_solution}

This section describes our solution to the problems P1 and P2. The method \emph{\textbf{A}ugmented \textbf{R}ay-\textbf{T}racing po\textbf{S}t-processing} (\acs{ARTS}) can be implemented in the \ac{WCG} module of a digital twin framework to solve P1.
After the discussion of \acs{ARTS}, methods to solve P2 are then presented.

\subsection{Augmented Ray-tracing post-processing (ARTS) method}
 
\begin{figure}[htp]
    \centering
    \includegraphics[scale=0.33]{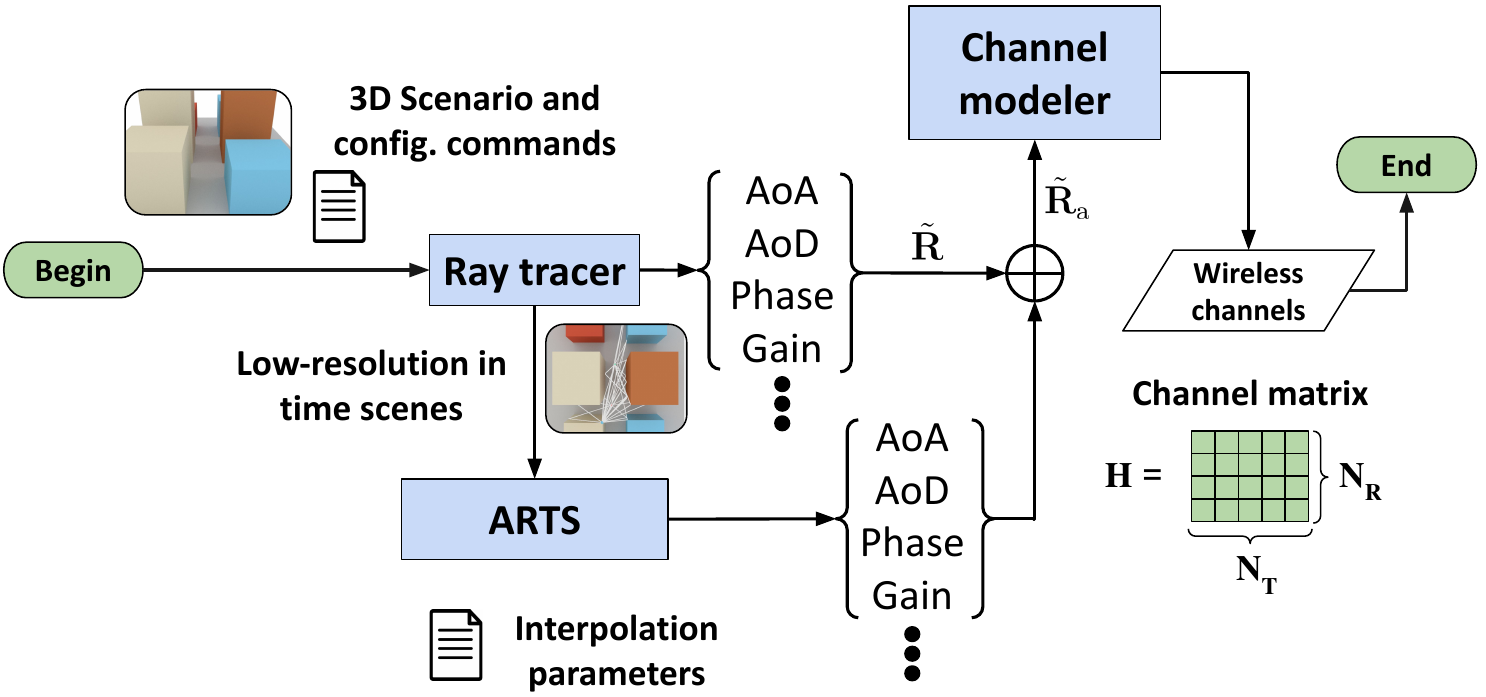}
    \caption{\ac{WCG} workflow considering \acs{ARTS} proposed method to speed up the \ac{RT} channel generation.}
    \label{fig:our_method}
\end{figure}

The input-output relationships in our method are described in Fig. \ref{fig:our_method}. The core of this implementation is represented by the \acs{ARTS} block and is based on three main steps, which are based on the concepts of pairing rays, equivalent rays, and stable segments. 

This comparison for pairing and checking ray equivalence relies on a corresponding face \ac{ID}. The faces of all objects composing the scene are uniquely identified, and the interaction position $(x,y,z)$ allows one to obtain the face \ac{ID} even in case the ray tracer does not directly provide it. Hence, the extended interactions set  $I_r^n$
 of ray $R_r^n$ is denoted as 
$\mathcal{I}_r^n = \{ (\mathbf{x}_r^n[1], t^n_r[1], d^n_r[1]), \ldots,  (\mathbf{x}_r^n[I_r^n], t^n_r[I_r^n], d^n_r[I_r^n]) \}$, where $d$ is an integer number. So, to achieve successful ray generation with small error, the ray augmenter (\acs{ARTS} block) leverages spatial consistency between scenes. In other words, before generating unknown ray parameters, each known ray—equivalent to another—is paired to form a stable segment. This pairing is essential because the ray tracer does not preserve the index of each ray throughout the simulation over time. The detailed structure of the \acs{ARTS} method is outlined in Algorithm~\ref{alg: wcg_our_method}.
\begin{algorithm}
\caption{ARTS algorithm}
\label{alg: wcg_our_method}
\begin{algorithmic}[1]
\State \textbf{Input:} $\Rseq$, $U$
\State \textbf{Output:} $\Rseq_\text{a}$
\State $M \gets N \times U - (U-1)$
\State $\Rseq_\text{a} \gets [\mathcal{R}'_1, \mathcal{R}'_2, \dots, \mathcal{R}'_p, \dots, \mathcal{R}'_M]$ \Comment{Initialize empty upsampled sequence}
\For{$n=1$ to $N-1$} \Comment{Pairing rays}
\For{$r=1$ to $K_n$}
\For{$j=1$ to $K_{n+1}$}
\If{$|\mathcal{I}_r^{n}| = |\mathcal{I}^{n+1}_j|$ and $d^n_r = d_{j}^{n+1}$}
\State $R_{\text{aux}} \gets R^{n+1}_r$
\State $R^{n+1}_r \gets R^{n+1}_j$
\State $R^{n+1}_j \gets R_{\text{aux}}$
\EndIf
\EndFor
\EndFor
\EndFor
\For{$n=1$ to $N-1$}\Comment{Iterate over known scenes}
    \State $\mathcal{R}'_n \gets \mathcal{R}_n$
    \For{$r=1$ to $K_n$} \Comment{Iterate over rays}
        \For{$p=n+1$ to $(n+U)-1$} 
            \If{ \text{is\_equivalent($R^n_r, R^{n+1}_r$)}}
            \State $\mathbf{p}^p_r \gets \mathbf{p}_r^n+\frac{p-1}{U}(\mathbf{p}^{n+1}_r-\mathbf{p}^n_r)$
            \Else
            \State \( B \sim \mathcal{B}(0.5) \)
                \If{$B = 0$}
                \State $\mathbf{p}^p_r \gets \mathbf{p}^n_r$ \Comment{Deals with dead ray}
                \Else
                \State $\mathbf{p}^p_r \gets \mathbf{p}^{n+1}_r$ \Comment{Deals with born ray}
                \EndIf
        \EndIf
        \EndFor
    \EndFor
\EndFor
\end{algorithmic}
\end{algorithm}

We now consider a simple example with two known scenes: the section of the algorithm related to pairing rays (lines 5-15) verifies if two rays achieve these conditions: (i) be consecutive in time (neighbors); (ii) have the same number of interactions, which means $|\mathcal{I}_r^n| = |\mathcal{I}_j^{n+1}|$; (iii) have the same face \ac{ID}. If so, we can pair two rays considering all interaction points, based on the comparison between the pair of object faces \ac{ID} of each interaction defined by the $d$. Once the pairing is done, we need to verify all equivalences to create a stable segment. This task is implemented in the \texttt{is\_equivalent} function (line 20). 

\begin{figure}[htp]
    \centering
    \includegraphics[scale=0.41]{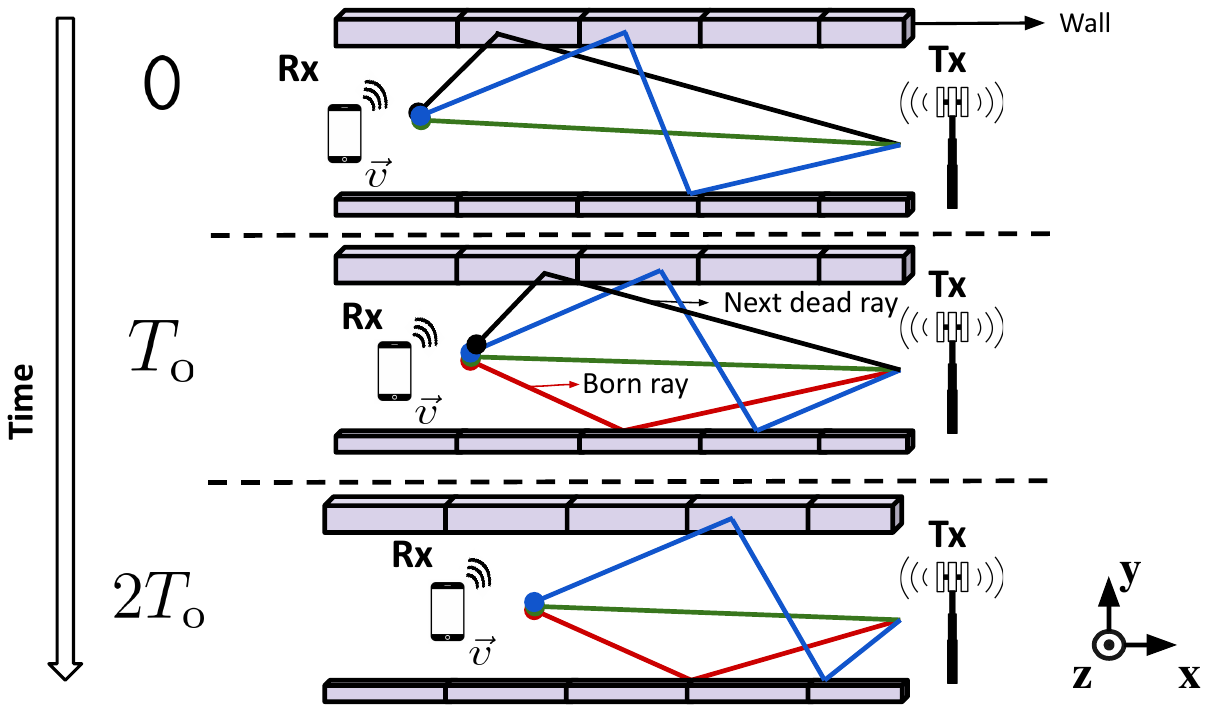}
    \caption{Representation of stable segment (in green and blue) between scenes in 0 and $2T_\textrm{o}$, in an environment with dead and born rays. We can assume that scene in $T_\textrm{o}$ should be interpolated.}
    \label{fig: born_and_dead_rays}
\end{figure}

An example of it can be visualized in Fig. \ref{fig: born_and_dead_rays}, hypothetically considering that scene 2 is unknown, and we would like to generate the \ac{LOS} ray (in green) in this scene. We will find that the \ac{LOS} ray in scenes 1 and 3 composes stable segments since all requirements are accomplished. With this equivalence verified, we can use a linear interpolation method (line 21) defined by 
\begin{equation}
\mathbf{p}^p_r = \mathbf{p}_r^n+\frac{p-1}{U}(\mathbf{p}^{n+1}_r-\mathbf{p}^n_r) ,
\end{equation}
\noindent where $1 < p < U+1 \hspace{0.15cm} \forall p \in \mathbb{Z}^+$.

When the equivalence is not detected between two rays (line 22) to form a stable segment (for instance the \ac{NLOS} red ray in Fig. \ref{fig: born_and_dead_rays}), this is depicted as a case of \textit{dead} or \textit{born} ray~\cite{Li2021}, which means that there exists an untraceable ray in the unknown scene that was born with a new characteristic or died in this exact scene, as depicted in Fig. \ref{fig: born_and_dead_rays}. Therefore, as it is not possible to know exactly if a ray will die or be born, we can model this behavior with different levels of complexity~\cite{Li2021}. In the \acs{ARTS} method, we adopt a simplified approach, considering that an untraceable ray can be modeled via a random variable $B$ obeying a Bernoulli distribution (line 23), $B \sim \mathcal{B}(\ell)$, with $\ell = 0.5$. So, given that a ray is untraceable, we have an equal probability of having a \textit{born} or \textit{dead} ray in the scene to be generated. 

Therefore, given an untraceable ray $R^p_r$ is detected in the $p$-th scene, if it was assigned a dead ray, this ray will assume characteristics from the $r$-th ray from the last known scene (line 25). Otherwise, if it was assigned as a born ray, this ray will assume the characteristics from the $r$-th ray from the next scene (line 27).

\subsection{Solutions to problem P2}
\begin{figure}[htp]
    \centering
    \includegraphics[scale=0.35]{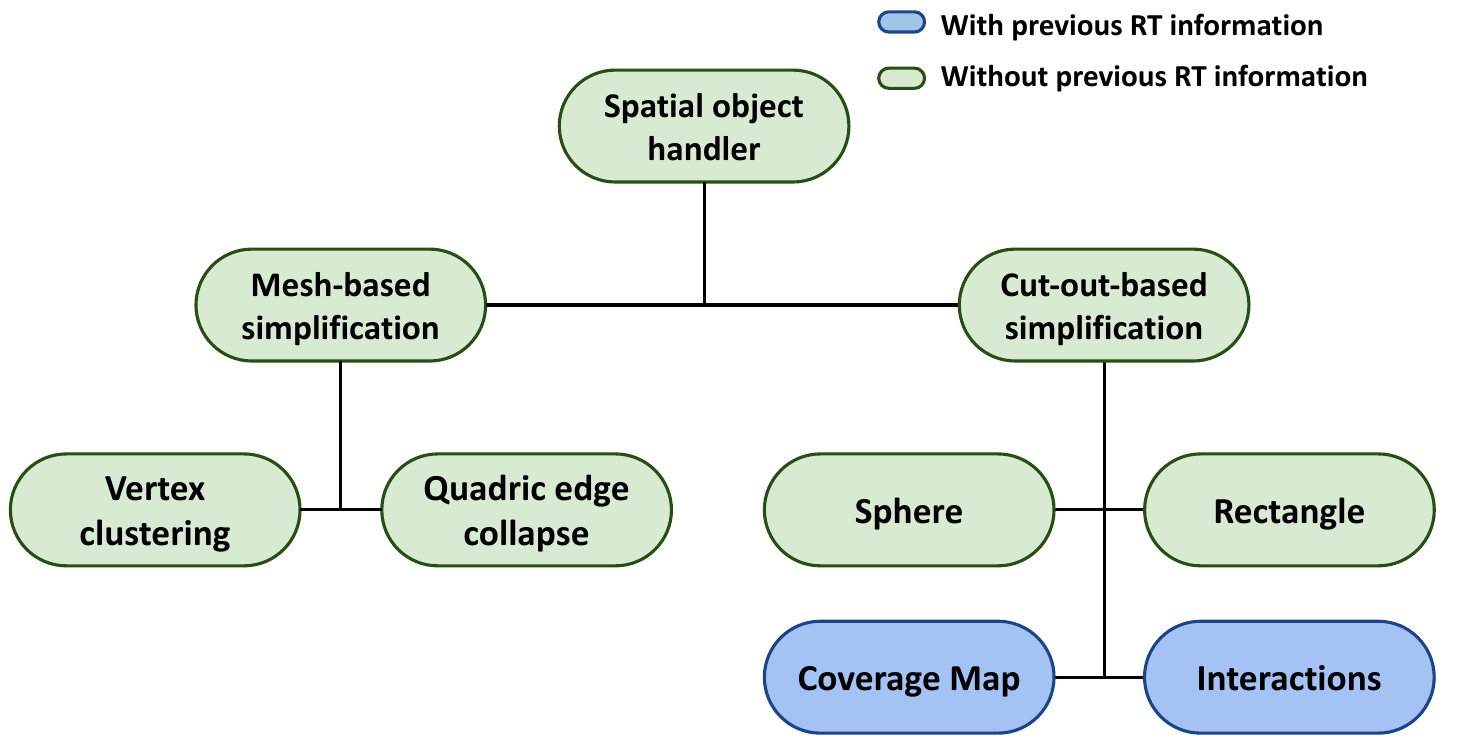}
    \caption{Taxonomy of the supported methods by the \ac{SOH} module.}
    \label{fig:3D_module}
\end{figure}

The next paragraphs discuss different solutions to Problem P2, including two classic \emph{mesh-based} methods (also referred to as \emph{decimation} algorithms) and the proposed \emph{cut-out-based simplification}. The mesh-based methods preserve the number of objects in a scene, while cut-out-based simplification discards objects that are not relevant to the \ac{RT}.
As indicated in Fig. \ref{fig:3D_module}, it is considered that a network digital twin \ac{SOH} module can execute methods from one or both submodules: mesh-based and cut-out-based simplification. 
The mesh-based methods represent the state-of-the-art in 3D scenario simplification and are implemented in software such as Blender~\cite{blender} and MeshLab~\cite{cignoni2008meshlab}.

\subsubsection{Mesh-based methods}

\begin{figure}[htp]
    \centering
    \includegraphics[scale=0.39]{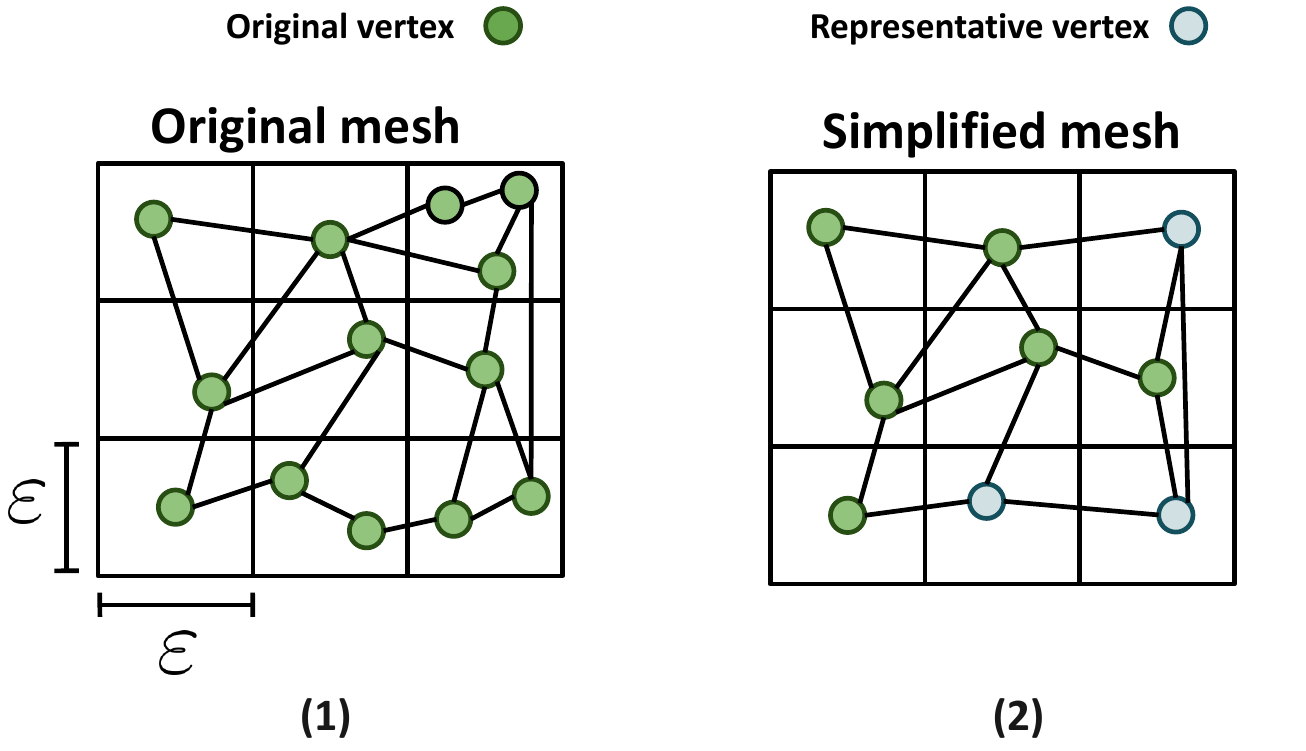}
    \caption{Vertex clustering algorithm operating in a 2D mesh in 3$\times$3 grid with a cell area of $\varepsilon^2$. Subfigure (1) represents the original mesh while (2) is the result based on the representative vertex approach.}
    \label{fig: vertex_clustering}
\end{figure}

In our \ac{SOH} module, two mesh-based algorithms are implemented: \emph{vertex clustering} and \emph{quadric edge collapse}~\cite{garland1997}. The vertex clustering algorithm operates on a grid of pixels (for 2D meshes) or voxels (for 3D meshes) of size \( V \). To illustrate this, consider a \(3 \times 3\) grid applied to a 2D mesh, as shown in Fig.~\ref{fig: vertex_clustering}(1). The same concept extends to 3D meshes without loss of generality. Given a cell size determined by \(\varepsilon\), which corresponds to an area of \(\varepsilon^2\) (11.1\% of the original pixel size), the algorithm aims to reduce the vertex count by clustering multiple vertices within each grid cell and assigning a representative vertex \( L \).  In other words, if a cell contains multiple vertices, they are mapped to a single representative vertex. This vertex is computed either as the weighted average of the coordinates of all vertices in the cell or as the position of the most heavily weighted vertex. The weights are determined based on two criteria~\cite{rossignac1993multi, rossignac1997geometric}: (i) the probability that a vertex lies on the object's silhouette from a given viewpoint, and (ii) the presence of a large face associated with the vertex. A vertex that satisfies both criteria—having a high silhouette probability and being part of a large face—is assigned a higher weight.  

Then, considering that we have $Q$ vertices inside a grid cell and the weighted average method, the representative vertex $L$ is given by
\begin{equation}
    L = \frac{1}{Q}\sum\limits_{i=1}^{Q} (Yv_{x_{i}}, Yv_{y_{i}}),
\end{equation}
\noindent where $Y$ is the weight of the vertex.
Hence, given the evaluated representative vertex, the simplified mesh looks like Fig. \ref{fig: vertex_clustering}(2).

\begin{figure}[htp]
    \centering
    \includegraphics[scale=0.4]{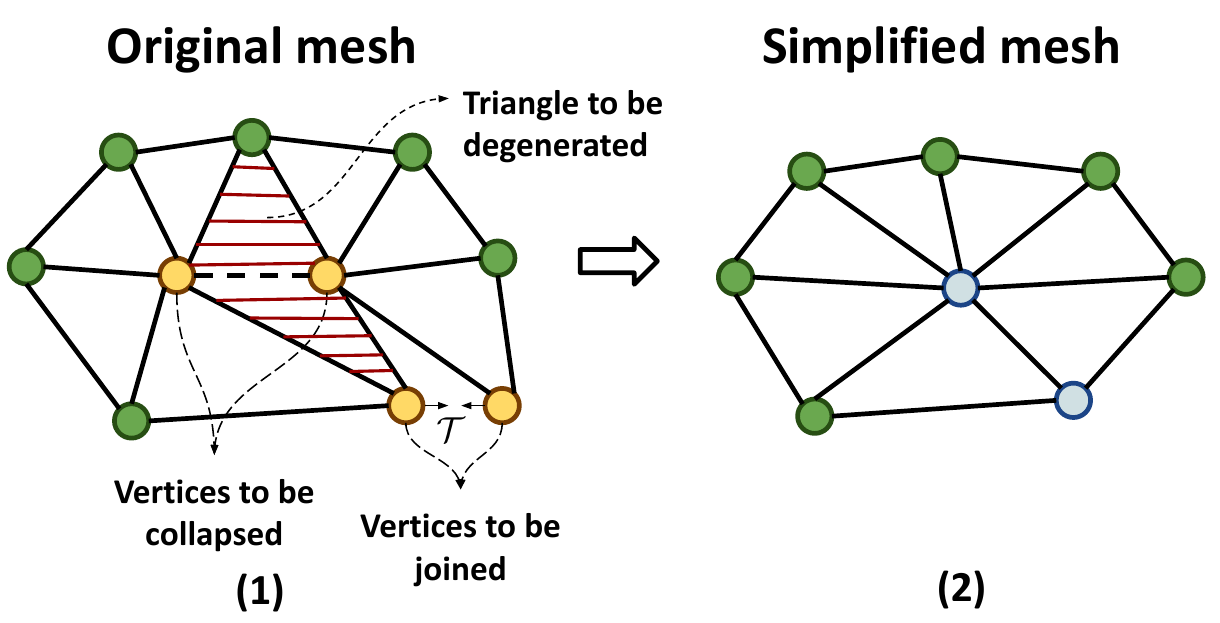}
\caption{Illustration of the quadric edge collapse algorithm applied to a 2D mesh. In (1), the dashed edge represents the edge selected for collapse, while the hashed triangles indicate the affected regions. The two arrows pointing toward each other signify the vertices that will be merged. In (2), the resulting 2D mesh is shown after the edge collapse operation.}
    \label{fig:quadric_edge_collapse}
\end{figure}

The quadric edge collapse method operates by iteratively contracting pairs of vertices considering a minimization of the quadratic error \cite{garland1997}. If an edge is selected for degeneration, all associated triangles are subsequently removed. However, if a pair of non-edge vertices is chosen, two separate regions of the mesh are merged into one. These two processes are illustrated in Fig.~\ref{fig:quadric_edge_collapse}, with step (1) depicting the contraction process and step (2) showing the resulting 2D mesh.  

This minimization process suggests an important geometric interpretation: if a vertex is associated with large faces, it will contribute significantly to the error metric. This makes intuitive sense, as such a vertex belongs to a relatively simple region of the 3D object where further simplification would lead to a substantial loss of geometric fidelity~\cite{garland1997}. Conversely, vertices linked to smaller faces contribute to a more intricate surface, offering greater flexibility in mesh simplification compared to objects composed predominantly of large faces.

The following paragraphs introduce the cut-out simplification methods proposed in this article. These methods can be used independently or in combination with mesh-based techniques to address Problem P2.

\subsubsection{CUT-OUT-BASED SIMPLIFICATION METHODS}

The right branch of Fig.~\ref{fig:3D_module} lists the proposed cut-out-based simplification methods, which aim to remove objects or regions from the original 3D scenario to accelerate conventional \ac{RT} by reducing the area that needs to be rendered. Unlike traditional simplification techniques that focus on preserving the 3D visual appearance, these methods prioritize maintaining the \ac{RF} behavior as accurately as possible.

A key innovation in this approach is scenario simplification driven by the goal of preserving \ac{RF} behavior, rather than using a predefined grid. The idea is to retain only the most relevant parts of the scenario—those that a ray is likely to reach. For example, if \acp{TRP} are positioned in the northwest of a city, an \ac{RT} simulation would reveal that no rays reach buildings in the distant southwest, spanning kilometers. This approach is conceptually similar to occlusion culling in 3D rendering, widely used in \textit{open world} games, where only the visible portions of large environments are loaded at runtime~\cite{Wu2022}.

To achieve this, four methods are considered:
\begin{itemize}
    \item 
 \emph{Sphere} Method: removes all parts of the scenario outside a sphere centered at the midpoint between a \ac{Tx} and \ac{Rx}, with a radius equal to the distance between them.
    \item 
 \emph{Rectangle} Method: removes areas outside a rectangular region defined by the \ac{Tx} and \ac{Rx}, with an additional margin along the $x$ and $y$ axes. This margin is determined based on the Tx-Rx distance and scaled by a constant factor.
    \item 
 \emph{Interaction} Method: requires at least one \ac{RT} simulation to identify and retain only the 3D objects surrounding interaction points, discarding unrelated elements. This ensures that only objects relevant to the wireless channel are preserved.
    \item 
 \emph{Coverage Map} Method: also requires an \ac{RT} simulation and removes regions based on a predefined power threshold (in \si{dB}). The scenario is pruned by discarding areas where the received power falls below a threshold, optimizing the input for subsequent \ac{RT} simulations.
\end{itemize}

These methods enable simplifying the 3D scenario while keeping key \ac{RF} characteristics.

\section{Evaluation} 
\label{sec: evaluation}
All experiments were performed using Sionna\textsuperscript{\texttrademark}~RT v$0.19.2$~\cite{sionna_rt_docs} in different computers. For evaluating solutions to P1 (\ac{WCG} methods), the server was equipped with an Intel\textsuperscript{\textregistered}~Core\textsuperscript{\texttrademark}~i7-10700F CPU @ 2.90 GHz, 128 GB RAM. For the P2 solutions (\ac{SOH} methods), the server was equipped with Intel\textsuperscript{\textregistered}~Xeon\textsuperscript{\texttrademark}~Gold 5419Y, 32 GB RAM.
In terms of channel models, we considered two types: a spatial 2D channel model and a 3D channel model~\cite{Heath2016,Trindade2018,Ruah2024}. The former model is defined by
\begin{equation}
    \label{eq: geometric_ula_channel}
    \mathbf{H}_n(f_\textrm{s}) = \sum\limits_{r=1}^{K_n}\alpha_r e^{-j\Omega_r}\mathbf{a}_\text{R}(\theta_{r}^\textrm{A})\mathbf{a}_\text{T}^*(\theta_{r}^\textrm{D}),
\end{equation}
\noindent where $\mathbf{H}_n$ is the channel matrix for the scene $n$, $\mathbf{a}_\text{T}$, $\mathbf{a}_\text{R}$ are the steering vectors at the transmitter (Tx) and receiver (Rx); $\alpha_r$, $\theta_{r}^\textrm{A}$, $\theta_{r}^\textrm{D}$ and $\Omega_r$ are listed in Table~\ref{tab:ray_parameters}. Moreover, the phase $\Omega_r$ is the most difficult parameter to predict among the ones listed in Table~\ref{tab:ray_parameters}.
It can be written as
\begin{equation}
\label{eq: linear_phase}
    \Omega_r = \psi_r -2\pi f_\textrm{s} \tau_r,
\end{equation}
where $\psi_r$ is the phase parcel that depends on the ray interactions and the electromagnetic properties (e.\,g., permittivity, conductivity, permeability, and scattering coefficient)
of the respective materials~\cite{Ruah2024}, $f_\textrm{s}$ is the subcarrier frequency, and $\tau_r$ is the $r$-th path delay as listed in Table~\ref{tab:ray_parameters}. 

Furthermore, the spatial 3D channel model is defined by
\begin{equation}
    \label{eq: geometric_upa_channel}
    \mathbf{H}_n(f_\textrm{s}) = \sum\limits_{r=1}^{K_n}\alpha_r e^{-j\Omega_r}\mathbf{a}_\text{R}(\theta_{r}^\textrm{A}, \phi_{r}^\textrm{A})\mathbf{a}_\text{T}^*(\theta_{r}^\textrm{D}, \phi_{r}^\textrm{D}),
\end{equation}
\noindent where $\phi_{r}^\textrm{A}$ and $\phi_{r}^\textrm{D}$ are also listed in Table \ref{tab:ray_parameters}.

Besides both \ac{WB} channel models, in the evaluation of our proposed methods for P1, we also consider their \ac{NB} variations, assuming a sufficiently small channel bandwidth $W$.

As the accuracy figure of merit $\overline{D}(\Rseq, \Rseqsimpler)$, we adopt the \ac{NMSE} between the corresponding channels, defined as
\begin{equation}
    \text{NMSE}_{\textrm{dB}} = 10\log_{10}\left\{\dfrac{||\textbf{H} - \hat{\textbf{H}}||^2_2}{||\textbf{H}||^2_2}\right\},
\end{equation}
\noindent where $\hat{\mathbf{H}}$ is the augmented \ac{MIMO} channel obtained from interpolated rays or the channel corresponding to the result of a 3D scene simplification method, and $\mathbf{H}$ is the ground truth channel created by the conventional ray tracer.

Finally, in order to evaluate the solutions to Problems P1 and P2, we consider each solution independently, in spite of the fact that in practice they can be combined by the \ac{WCG} and \ac{SOH} modules.

\subsection{RT augmentation (P1)} 

To evaluate the solutions of problem P1, we considered the channel characteristics in both \ac{NB} ($\tau_r W \ll 1$) and \ac{WB} (which we adopted $W = 100$ \si{MHz} and 64 subcarriers) regimes, based on Eq. (\ref{eq: geometric_ula_channel}) and Eq. (\ref{eq: geometric_upa_channel}). Specifically, two antenna array configurations were employed: (i) a \ac{MIMO} system with a \ac{ULA} comprising eight transmit and four receive antennas, and (ii) a \ac{MIMO} system with a \ac{URA}
featuring a $4 \times 4$ antenna layout in the \ac{Tx} and $8 \times 4$ in the \ac{Rx}. The carrier frequency $f=2.14$~\si{GHz}, an isotropic radiation pattern and cross-polarization were assumed in all configurations.

\begin{table}[htp]
    \centering
    \caption{\ac{RT} simulation parameters used to evaluate the proposed solution for the problem P1}
    \scalebox{1}{\begin{tabular}{cc}
    \toprule
      Simulation parameter & Parameter\\
    \midrule
       Carrier frequency (\si{GHz})  & $2.14$ \\
       Number of Tx & $1$\\
       Number of Rx & $1$\\
       Radiation pattern & Isotropic \\
       Polarization & Cross polarization\\
       Max. number of interactions per ray & $3$\\
       \ac{WB} channel bandwidth (\si{MHz}) & 100  \\
\bottomrule
    \end{tabular}}
\label{tab:wcg_simulation_parameters}
\end{table}

We compare \acs{ARTS} with a baseline method that increases the number of channel samples by matrices $\mathbf{H}$ interpolation.
In this case, all operations occur in the ``matrix domain'', instead of the ``rays domain'' used by \acs{ARTS}. For instance, considering a simple example with two scenes $\mathcal{R}_1$ and $\mathcal{R}_2$, this baseline method only needs to generate the channels $\mathbf{H}_1$ and $\mathbf{H}_2$, respectively, and obtain extra channels using interpolation as in
\begin{equation}
    \label{eq: 1d_linear_matrix_interpolation}
    \hat{\mathbf{H}}_p = \mathbf{H}_1 + \dfrac{p-1}{U}(\mathbf{H}_2 - \mathbf{H}_1),
\end{equation}
\noindent where $1 < p < U+1 \hspace{0.15cm} \forall p \in \mathbb{Z}^+$.

To compare the performance of the \acs{ARTS} method with this baseline in solving P1, we used three different urban macro scenarios: \textit{Munich}, \textit{Etoile}, and \textit{St. Canyon}~\cite{sionna_rt_docs}. For all of them, the \ac{Tx} height is greater than 25~\si{m}. In terms of mobility, we have a single mobility condition, where only the \ac{Rx} moves along a linear trajectory. For all \ac{Rx}, we consider a constant velocity of 3 \si{km/h}. The simulations consider only reflection and diffraction, and three different upsampling factors: $U = 2$, $U = 4$, and $U = 10$. The value $U=2$ is the smallest upsampling factor, while $U=10$ is much more challenging. Given $U$, the sampling interval $T_o$, and the receiver speed, one can estimate the number of meters the receiver moved between neighboring scenes.

The conventional \ac{RT} was executed for a total of $1000$ scenes, which are then processed and split into
 ``known'' channels and the ones that play the role of ground truth to assess augmentation.
Table \ref{tab: wcg_experiments} summarizes the experiments.

\begin{table}[htp]
    \centering
    \caption{Characterization of experiments performed to evaluate the proposed \acs{ARTS} method}
    \scalebox{0.82}{\begin{tabular}{cccccccc}
    \toprule
       \makecell{\#} & \makecell{Array \\ pattern}  & $U$  & \makecell{\# Known \\scenes} & \makecell{Known scenes\\ density)} & $T_\textrm{o}$ & $T_\textrm{s}$ & \makecell{Channel\\type}\\
    \midrule
         
        $1$ & ULA & $2$ & $500$ & $1$ scene every $0.4$ \si{m} & $0.48$ & $0.24$ & NB \\
        $2$ & URA & $2$ & $500$ & $1$ scene every $0.4$ \si{m} & $0.48$ & $0.24$ & NB \\
        $3$ & ULA & $10$ & $100$ & $1$ scene every $2$ \si{m} & $2.4$ & $0.24$ & NB \\
        $4$ & ULA & $4$ & $250$ & $1$ scene every $0.8$ \si{m} & $0.96$ & $0.24$ & WB \\
    \bottomrule
    \end{tabular}}
    \label{tab: wcg_experiments}
\end{table}

\begin{table*}[htbp]
    \centering
    \caption{Comparison of results with \acs{ARTS} and the baseline for all four experiments}
    \scalebox{1.3}{\begin{tabular}{cc|cc|cc|cc}
    \toprule
    \multirow{4}{*}{Experiments} & \multicolumn{6}{c}{\ac{NMSE} (\si{dB})}& \\ 
    \cmidrule{2-8}
    & &
    \multicolumn{2}{c|}{Munich}& \multicolumn{2}{c|}{Etoile} & \multicolumn{2}{c}{St. Canyon}\\ \cmidrule{2-8}
    &Method & Average& Max.& Average& Max.& Average & Max.\\ \midrule
    \multirow{2}{*}{Experiment 1} &  Matrix linear interp.&$-1.40$&$5.56$&$-8.53$&$\mathbf{3.93}$&$-14.15$&$4.62$\\ 
    & Proposed ARTS method&$\mathbf{-56.38}$&$\mathbf{4.44}$&$\mathbf{-67.26}$&$4.18$&$\mathbf{-75.91}$&$\mathbf{4.55}$\\ \midrule
    \multirow{2}{*}{Experiment 2} & Matrix linear interp. &$1.72$&$5.10$&$-1.85$&$4.41$&$-2.95$ & $9.15$\\
    &Proposed ARTS method&$\mathbf{-55.39}$&$\mathbf{2.85}$& $\mathbf{-65.33}$ &$\mathbf{4.36}$&$\mathbf{-70.10}$ & $\mathbf{3.78}$\\
    \midrule
    \multirow{2}{*}{Experiment 3} & Matrix linear interp. &$-1.33$&$\mathbf{6.11}$&$-2.46$&$5.95$&$-5.16$ & $11.31$\\
    &Proposed ARTS method&$\mathbf{-18.43}$&$6.27$&$\mathbf{-29.57}$&$\mathbf{4.65}$&$\mathbf{-34.18}$ & $\mathbf{7.32}$\\
    \midrule
    \multirow{2}{*}{Experiment 4} & Matrix linear interp. &$-0.43$&$\mathbf{5.94}$&$-5.74$&$\mathbf{4.62}$&$-9.81$ & $\mathbf{7.73}$\\
    &Proposed ARTS method&$\mathbf{-56.47}$&$9.34$&$\mathbf{-67.63}$&$5.22$&$\mathbf{-70.35}$ & $14.79$\\
    \bottomrule
    \end{tabular}}
    \label{tab: experiments_aggregation_measures}
\end{table*}

\begin{figure}[htp]
    \centering
    \includegraphics[scale=0.5]{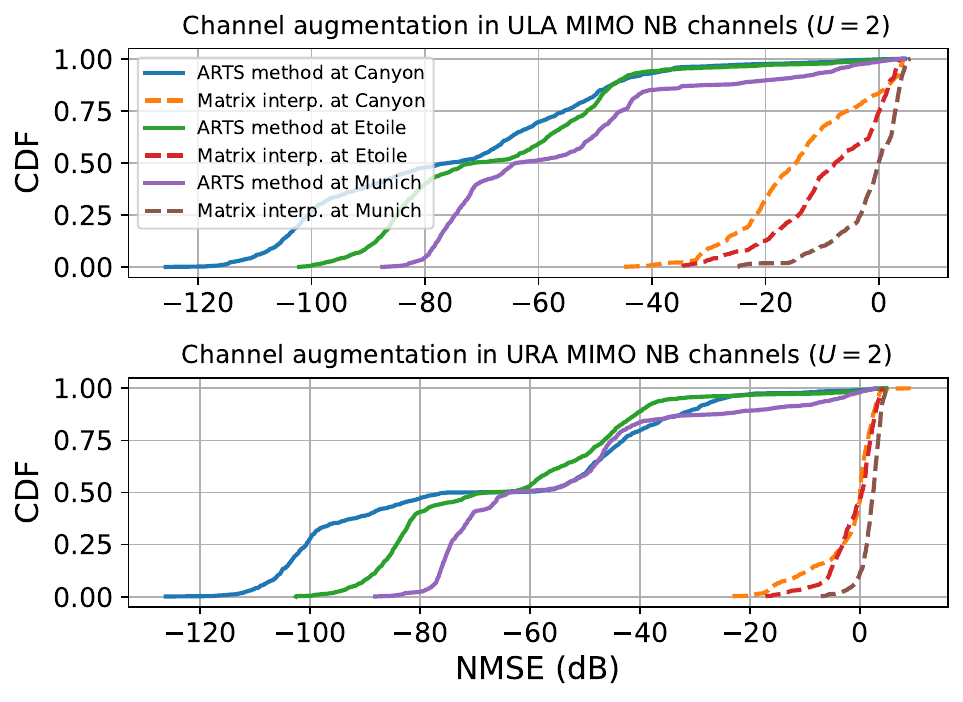}
    \caption{\ac{CDF} of \ac{NMSE} for channel augmentation in \ac{ULA} and \ac{URA} \ac{MIMO} channels, considering our proposed method for Problem P1 and matrix interpolation at Munich, Etoile, and St. Canyon scenario.}
    \label{fig: wcg_experiment_1_nmse}
\end{figure}

In the Experiments 1 and 2, $500$ channels are known, provided by the ray tracer, and augmentation is used to estimate the remaining $500$ \ac{NB} channels, considering \ac{ULA} and \ac{URA} array patterns. The results are presented in Fig. \ref{fig: wcg_experiment_1_nmse}, where the \acp{CDF} shows that \acs{ARTS} outperforms interpolation in the matrix domain. 

\begin{figure}[htp]
    \centering
    \includegraphics[scale=0.47]{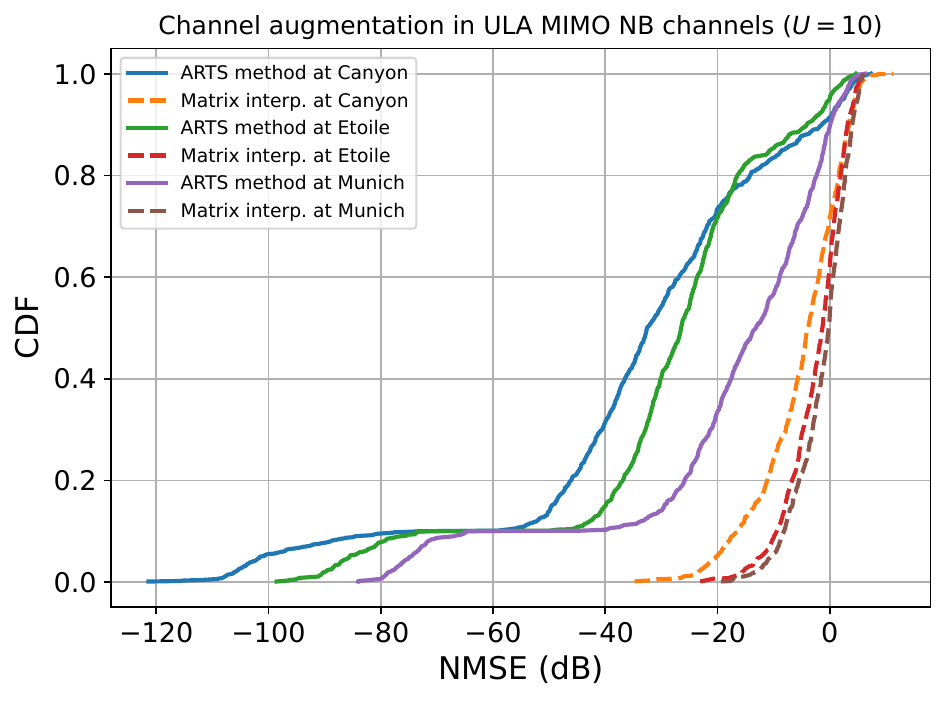}
    \caption{\ac{CDF} of \ac{NMSE} for channel augmentation in \ac{ULA} \ac{MIMO} channels considering our proposed method and matrix interpolation, with a fewer number of known samples.}
    \label{fig: wcg_experiment_3_nmse}
\end{figure}

As mentioned, Experiment 3 is more challenging given that $U=10$. Specifically, it required generating $900$ scenes from information provided by $100$ known scenes.
The results in Fig. \ref{fig: wcg_experiment_3_nmse} illustrate the performance under this configuration. 

The results in Fig. \ref{fig: wcg_experiment_3_nmse} indicate the expected decline in performance with respect to Fig. \ref{fig: wcg_experiment_1_nmse}. This degradation can be attributed to the lack of spatial consistency between the \ac{RT} samples. Specifically, the probability of obtaining two equivalent rays separated by 2 \si{m} is significantly lower compared to when the samples are only $0.4$ \si{m} apart. Despite this decline in \ac{NMSE}, the proposed \acs{ARTS} method 
outperforms the matrix interpolation method in all three scenarios, which is summarized in Table \ref{tab: experiments_aggregation_measures}.
\acs{ARTS} achieves a reduction of more than 17~\si{dB} in average \ac{NMSE}, while decreasing the maximum distortion in half of the cases.

\begin{figure}[htp]
    \centering
    \includegraphics[scale=0.47]{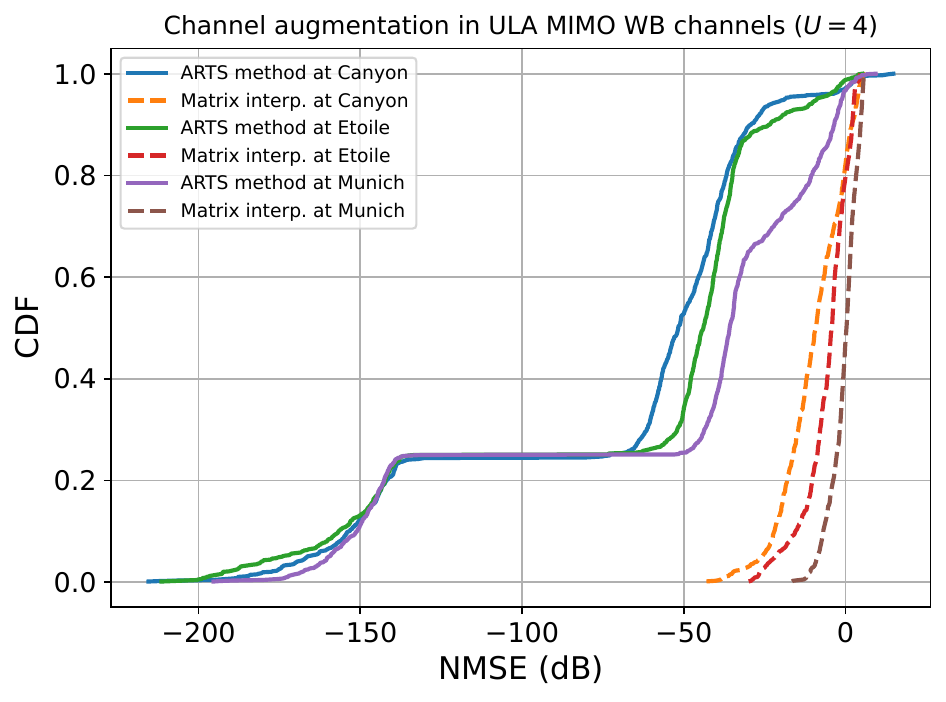}
    \caption{\ac{CDF} of \ac{NMSE} for channel augmentation in \ac{ULA} \ac{WB} \ac{MIMO} channels, evaluated under the same scenarios as the previous three experiments.}
    \label{fig:wcg_experiment_4_nmse}
\end{figure}

Finally, the last set of experiments aimed to evaluate our method under a \ac{WB} channel, which exhibits different characteristics from channel used in the previous three experiments, although conducted in the same scenarios. In this case, we employed an upsampling factor of $U = 4$. The results, shown in Fig.~\ref{fig:wcg_experiment_4_nmse}, demonstrate that the \acs{ARTS} method consistently outperforms the baseline under the \ac{WB} channel model.

\subsection{3D scenario simplification (P2)}

To validate the proposed solutions for the P2 problem, we consider only a \ac{ULA} \ac{MIMO} system, using the same number of \ac{Tx} and \ac{Rx} antennas as in P1, and adopt the \ac{NB} scenario derived from Eq.~\ref{eq: geometric_ula_channel}. The carrier frequency, radiation pattern, and polarization remain unchanged. An important difference in the \ac{RT} simulation, however, is that while the P1 solutions accounted only for reflection and diffraction, the P2 scenario also incorporates diffuse scattering. This inclusion increases the number of detected paths and, consequently, the complexity of the \ac{RT} process, resulting in more pronounced effects on both \ac{NMSE} and \ac{RT} simulation time.

Using this channel configuration, we defined two experiments considering one more metric: the \ac{RT} simulation time in comparison with that obtained in the original scenario. So, in the first experiment, we separately assessed all 3D simplification methods. In the second experiment, we evaluated the performance taking into account all possible combinations of mesh and cut-out-based simplification approaches. For both experiments, we considered the Modern City scenario, which was depicted in Fig. \ref{fig:original_scenario}. Modern City has a larger number of meshes when compared to the three scenarios (Munich, Etoile, and St. Canyon) used to validate the \ac{WCG} methods. More specifically, Modern City has 5\,922\,924 faces; while Munich, Etoile, and St. Canyon have 38\,938, 13\,058, and 74 faces, respectively. Furthermore, the scope of the two new experiments comprises six simplification methods, all defined in Fig. \ref{fig:3D_module}. Specific parameters for each method are  summarized in Table \ref{tab:3d_object_parameter_experiments}.

\begin{table}[htp]
    \centering
    \caption{Parameter values for all simplification methods used to assess P2 solutions}
    \scalebox{1}{\begin{tabular}{cc}
    \toprule
      Simplification method & Parameter value
      \\
      \midrule
        
       Quadric edge collapse  & $\lambda = 30$\%\\
       Vertex clustering & $\mathcal{J} = 5$\% \\
        
       Interactions cut-out & $C = 2$ \si{m} \\
       Coverage Map cut-out & $\Sigma = -140$ \si{dB} \\
       
       Sphere cut-out & $G = 50.94$ \si{m} \\
       Rectangle cut-out & $\mathcal{A} = 25.47$ \si{m} \\
    \bottomrule
    \end{tabular}}
    \label{tab:3d_object_parameter_experiments}
\end{table}

\begin{table*}[htbp]
    \centering
    \caption{Results obtained in terms of \ac{RT} duration and \ac{NMSE} for each simplification method in comparison with the original scenario}
    \scalebox{1}{\begin{tabular}{c|ccc|ccc}
    \toprule 
   \multirow{3}{*}{Environments} & 
    \multicolumn{3}{c|}{Linear track}& \multicolumn{3}{c}{Square track} \\ \cmidrule{2-7}
    & RT duration (min) & Average NMSE (\si{dB}) & SD NMSE & RT duration (min) & Average NMSE (\si{dB}) & SD NMSE\\
    \midrule
    Without simplification & $94.89$ &---&--- & $225.36$ & ---&---\\
    Coverage Map cut-out & $57.29$ & $\mathbf{-13.48}$&$11.38$& $136.50$ & $\mathbf{-7.80}$&$13.58$\\
    Sphere cut-out & $42.18$ & $-12.40$&$11.38$ & $76.10$ & $-4.09$&$9.91$\\
    Rectangle cut-out & $39.91$ & $-10.55$&$9.91$& $\mathbf{74.97}$ & $-2.45$&$10.00$\\
    Interactions cut-out & $\mathbf{39.26}$ & $-5.52$&$9.95$ & $93.81$ & $-5.11$&$10.96$\\
    Vertex clustering & $77.64$ & $-3.76$&$7.96$ & $188.80$ & $-1.02$&$8.76$\\
    Quadric edge collapse & $77.62$ & $-5.37$&$8.95$ & $188.15$ &$-2.23$&$10.60$\\
    \bottomrule
    \end{tabular}}
    \label{tab:duration_nmse_results}
\end{table*}

For the quadric edge collapse, we set the final percentage of preserved faces to \(\lambda = 30\%\). In the vertex clustering simplification approach, the voxel size \(\mathcal{J}\) was set to \(5\%\) of the original object volume.  
In the last four methods, the Interaction cut-out was configured with a threshold of \(C = 2\)~\si{m} along all three 3D axes, defining a cube with dimensions \(4 \times 4 \times 4\)~\si{m}, centered at the interaction point, where only objects within this region were retained. For the coverage map method, a power threshold of \(-140\)~\si{dB} was applied to remove all objects below this coverage level. Regarding the Sphere method, a sphere of radius \(G = 50.94\)~\si{m} was centered at the midpoint between the \ac{Tx} and \ac{Rx}, eliminating all objects outside this boundary. Finally, in the Rectangle method, a margin distance of \(\mathcal{A} = 25.47\)~\si{m} relative to the \ac{Tx} and \ac{Rx} was used to define the rectangular section, employing a constant factor of 2 to exclude all objects beyond this section.  

\begin{figure}[htp]
    \centering
    \includegraphics[scale=0.35]{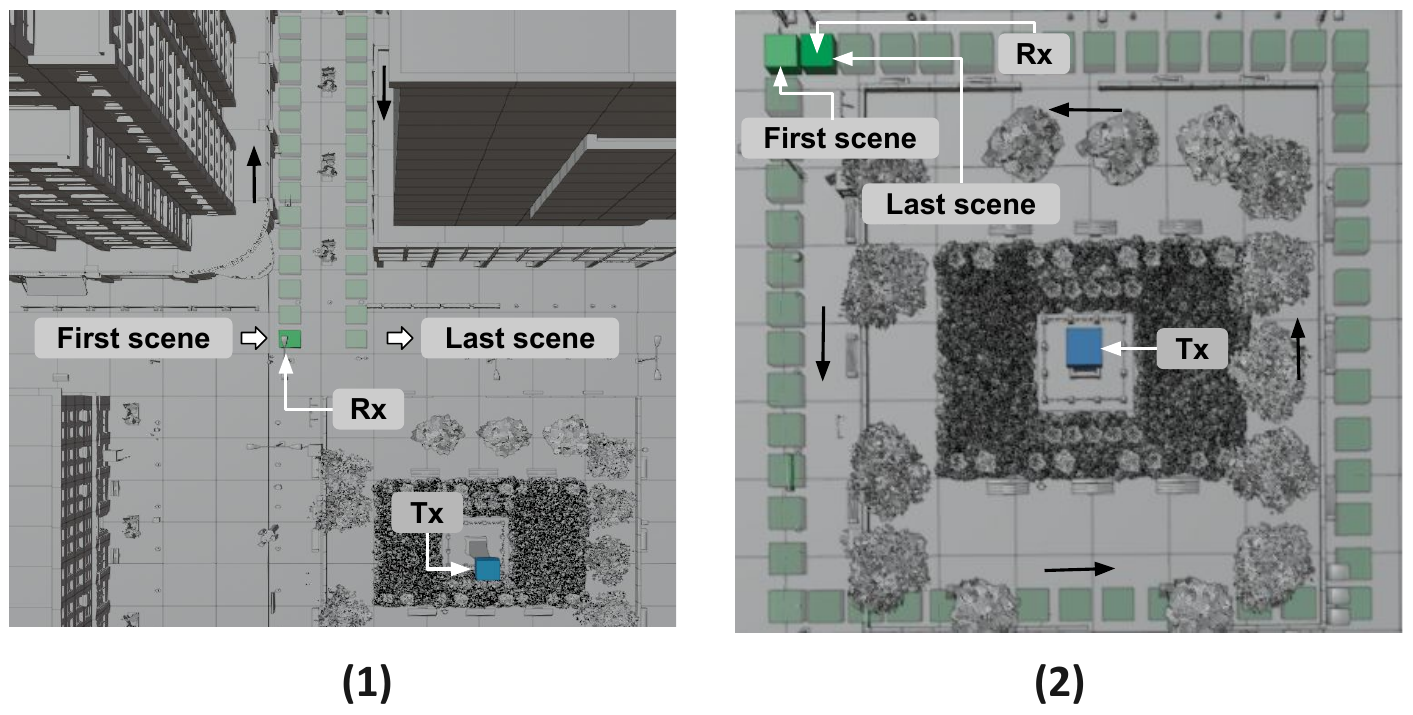}
    \caption{Two types of track mobility pattern adopted in the proposed solutions for P2 of the first experiment: linear (1) and square (2) track.}
    \label{fig:rx_path}
\end{figure}

To compute the \ac{NMSE} and the \ac{RT} runtime, for each simplification method, we considered two different track patterns: linear and square. For the linear track, we ran 162 scenes, varying the \ac{Rx}. For the square track pattern, we considered a higher scene density, with a total of 284 scenes. Both tracks are shown in Fig. \ref{fig:rx_path}. For both cases, after that, we took the average, and the \ac{SD} between the \ac{NMSE} values obtained. Thus, for this first experiment regarding P2, we obtained the results summarized in Table \ref{tab:duration_nmse_results}, divided by each track.

As the results show in Table \ref{tab:duration_nmse_results}, among all simplification methods, those that presented the best results with respect to the trade-off between \ac{RT} duration and the average \ac{NMSE} are related to the Coverage Map and Sphere cut-out methods. The former obtained a better performance at maintaining channel characteristics, with a \ac{NMSE} close to $-13.5$ and $-7.8$, considering linear and square track mobility patterns, respectively. However, this method had a longer duration than other cut-out methods, such as the Sphere cut-out. The latter method obtained a slightly worse \ac{NMSE}, equal to $-12.4$ in the linear track experiment, and $-4.09$ on the square track, but with a \ac{RT} simulation time less than the former method. Among the different cut-out methods, the best \ac{RT} runtime reduction was achieved using the Interactions cut-out on the linear track, with a time reduction of over 50\%. However, this came at the cost of significant oversimplification, leading to a larger discrepancy between the channel characteristics of the original and simplified scenarios, as this method removes nearly the entire environment. A similar trend was observed with the Rectangle method on the square track, which also reduced \ac{RT} duration but resulted in a less favorable \ac{NMSE}. Conversely, on the linear track, the Rectangle method provided a substantial \ac{RT} runtime reduction—close to that of the Interactions method—while offering a more balanced trade-off between complexity and accuracy.

Concerning the mesh simplification algorithms, i.e., the vertex clustering and the quadric edge collapse, both performed worse than the cut-out-based ones in terms of \ac{NMSE} and \ac{RT} duration. Both methods produced similar results, with the quadric edge collapse achieving a slightly better \ac{RT} duration and average \ac{NMSE} in both track patterns.

\begin{figure}[htp]
     \centering
     \includegraphics[scale=0.43]{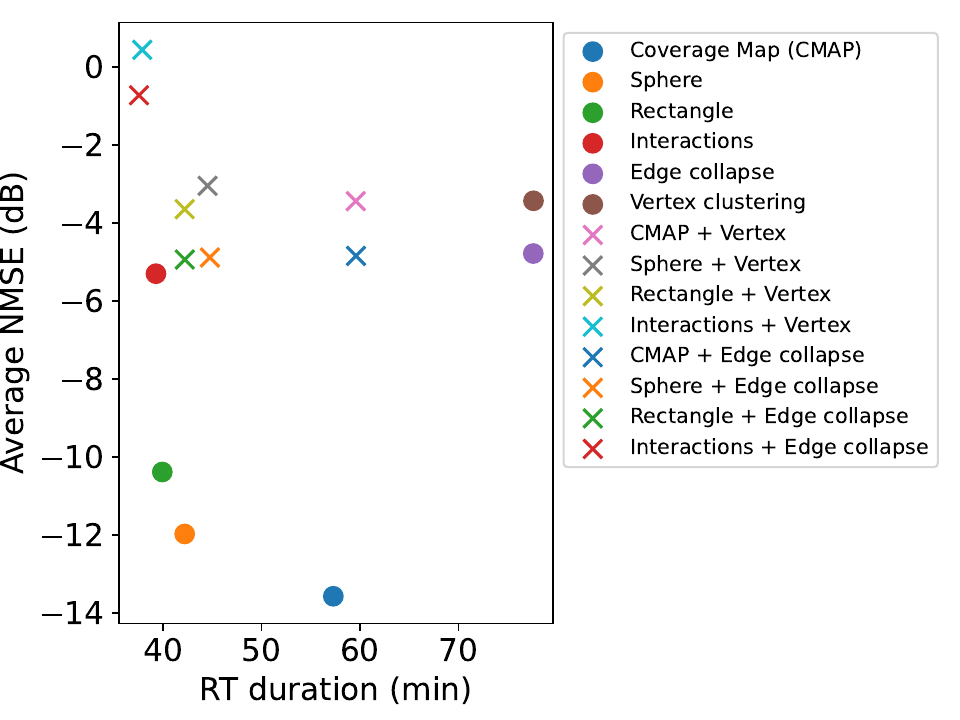}
     \caption{Tradeoff between \ac{NMSE} and \ac{RT} simulation time combining different P2 solutions considering a linear track pattern.}
     \label{fig:tradeoff_combination}
 \end{figure}

In the second experiment regarding Problem P2, using only the environment with a linear track, the eight possible combinations of both simplification method approaches, along with the metric from the first experiment, produced the scatter plot that depicts solutions in a multi-objective Pareto front as shown in Fig. \ref{fig:tradeoff_combination}.
This figure illustrates the tradeoff between \ac{RT} simulation time (in minutes) on the $x$-axis and the average \ac{NMSE} across all \ac{RT} scenes on the $y$-axis. These results provide insight into the impact of combining different simplification methods, and are  detailed in Table \ref{tab:duration_nmse_comb_results}. For instance, if a mesh simplification method is to be used, pairing it with the Coverage Map cut-out proves particularly effective, as it reduces simulation time while maintaining the \ac{NMSE}. However, it is important to apply these combinations with caution, as they can lead to increased distortion.

\begin{table}[htp]
\centering
\caption{Results obtained in terms of \ac{RT} duration and \ac{NMSE} for different combination of 3D simplification methods with \ac{Rx} mobility assuming a linear track}
\scalebox{0.85}{\begin{tabular}{c|cccc}
\toprule
\multicolumn{1}{c|}{\multirow{2}{*}{Environment}}     & \multicolumn{2}{c|}{\multirow{2}{*}{\makecell{Total \\ duration (min)}}} & \multicolumn{2}{c}{\ac{NMSE}}\\ 
\multicolumn{1}{c|}{}                      & \multicolumn{2}{c|}{} & \multicolumn{1}{c|}{Average}            & \multicolumn{1}{c}{\ac{SD}}\\ \midrule
\multicolumn{1}{c|}{Without simplification}     & \multicolumn{2}{c|}{$94.89$} & \multicolumn{1}{c|}{---}             & \multicolumn{1}{c}{---}\\
\multicolumn{1}{c|}{Coverage Map cut-out + Vertex clustering}  & \multicolumn{2}{c|}{$59.57$} & \multicolumn{1}{c|}{$-3.44$} & \multicolumn{1}{c}{$8.51$}\\
\multicolumn{1}{c|}{Sphere cut-out + Vertex clustering}        & \multicolumn{2}{c|}{$44.18$} & \multicolumn{1}{c|}{$-3.04$}          & \multicolumn{1}{c}{$7.83$}\\
\multicolumn{1}{c|}{Rectangle cut-out + Vertex clustering}     & \multicolumn{2}{c|}{$42.17$} & \multicolumn{1}{c|}{$-3.64$}          & \multicolumn{1}{c}{$8.38$}\\
\multicolumn{1}{c|}{Interactions cut-out + Vertex clustering}  & \multicolumn{2}{c|}{$37.87$} & \multicolumn{1}{c|}{$0.43$}           & \multicolumn{1}{c}{$8.55$}\\
\multicolumn{1}{c|}{Coverage Map cut-out + Edge collapse}     & \multicolumn{2}{c|}{$59.59$} & \multicolumn{1}{c|}{$-4.84$}           & \multicolumn{1}{c}{$8.87$}\\
\multicolumn{1}{c|}{Sphere cut-out + Edge collapse}     & \multicolumn{2}{c|}{$44.74$} & \multicolumn{1}{c|}{$-4.88$}           & \multicolumn{1}{c}{$8.84$}\\
\multicolumn{1}{c|}{Rectangle cut-out + Edge collapse}     & \multicolumn{2}{c|}{$42.19$} & \multicolumn{1}{c|}{$\mathbf{-4.93}$}           & \multicolumn{1}{c}{$8.25$}\\
\multicolumn{1}{c|}{Interactions + Edge collapse} & \multicolumn{2}{c|}{$\mathbf{37.54}$} & \multicolumn{1}{c|}{$-0.72$}           & \multicolumn{1}{c}{$9.51$}\\ \bottomrule
\end{tabular}}
\label{tab:duration_nmse_comb_results}
\end{table}

\section{Conclusions and Future Works}
\label{sec: conclusion}

This article posed two problems related to channel generation with \ac{RT} and a methodology to assess their solutions. This methodology relies on constructing ``ground truth'' results, which allows evaluations using objective figures of merit such as \ac{NMSE}. It was also explained how the two problems map to modules in an enabler tool for wireless channel modeling in an \ac{RT}-based \ac{NDT} system.\footnote{Source code will be available at: \url{https://github.com/lasseufpa/speedup-rt}}

The first problem (P1) regards augmenting an existing set of \ac{RT} results, to obtain a new set with improved time resolution via post-processing. The proposed method, called \acs{ARTS} outperformed a baseline that relies on 
interpolation in the ``matrix domain''. 
The \acs{ARTS} method explores spatial consistency between scenes and is capable of significantly speeding up \ac{RT}-based wireless channel generation.
For the tested scenarios, it is possible to decrease at least by half (using an upsampling factor $U=2$) the \ac{RT} simulation time, maintaining reasonable accuracy in channel generation.

This work also proposed 
different methods to address 3D scenario simplification (P2). Solutions to P2 are crucial to make feasible running current \ac{RT} software for  scenarios that represent large areas and/or have a large number of meshes. As solutions to P2, we proposed four different cut-out based methods that discard unnecessary regions, aiming that they do not impact the corresponding wireless channel. These methods  explore the trade-off between the realism of the generated channel and the time to process a 3D scenario. Our cut-out based method results demonstrated that it is possible to decrease the level of unnecessary details and reduce the simulation time by half, while retaining an accurate representation of the channel with respect to the ground truth.

Finally, for future work, we will provide results obtained by combining solutions to both problems P1 and P2. When these solutions operate together in modules such as \ac{WCG} and \ac{SOH}, one can achieve almost real-time execution for relatively complex 3D scenarios in \acp{NDT} based on \ac{RT}. 
We will also analyze the impact of the proposed methods on various communication metrics, particularly link-level indicators such as signal-to-noise ratio and channel capacity. Another ongoing work is the integration of the proposed methods into the 
CAVIAR 
framework~\cite{Borges2024}.
Efficient \ac{RT} pre and post-processing methods enable CAVIAR to execute system-level simulations using realistic site-specific wireless channels, and evaluations based on key performance indicators such as delay, jitter, and packet loss.


%


\ifCLASSOPTIONcaptionsoff
  \newpage
\fi



\bibliographystyle{IEEEtran}
\bibliography{bibtex/main}

\begin{thebibliography}{10}
\providecommand{\url}[1]{#1}
\csname url@samestyle\endcsname
\providecommand{\newblock}{\relax}
\providecommand{\bibinfo}[2]{#2}
\providecommand{\BIBentrySTDinterwordspacing}{\spaceskip=0pt\relax}
\providecommand{\BIBentryALTinterwordstretchfactor}{4}
\providecommand{\BIBentryALTinterwordspacing}{\spaceskip=\fontdimen2\font plus
\BIBentryALTinterwordstretchfactor\fontdimen3\font minus \fontdimen4\font\relax}
\providecommand{\BIBforeignlanguage}[2]{{%
\expandafter\ifx\csname l@#1\endcsname\relax
\typeout{** WARNING: IEEEtran.bst: No hyphenation pattern has been}%
\typeout{** loaded for the language `#1'. Using the pattern for}%
\typeout{** the default language instead.}%
\else
\language=\csname l@#1\endcsname
\fi
#2}}
\providecommand{\BIBdecl}{\relax}
\BIBdecl

\bibitem{almers2007}
\BIBentryALTinterwordspacing
P.~Almers, E.~Bonek, A.~Burr, N.~Czink, M.~Debbah, V.~Degli-Esposti, H.~Hofstetter, P.~Ky\"{o}sti, D.~Laurenson, G.~Matz, A.~F. Molisch, C.~Oestges, and H.~\"{O}zcelik, ``{Survey of Channel and Radio Propagation Models for Wireless {MIMO} Systems},'' \emph{EURASIP J. Wirel. Commun. Netw.}, vol. 2007, no.~1, p.~56, Jan. 2007. [Online]. Available: \url{https://doi.org/10.1155/2007/19070}
\BIBentrySTDinterwordspacing

\bibitem{Danping2019}
D.~He, B.~Ai, K.~Guan, L.~Wang, Z.~Zhong, and T.~Kürner, ``{The Design and Applications of High-Performance Ray-Tracing Simulation Platform for {5G} and Beyond Wireless Communications: A Tutorial},'' \emph{IEEE Communications Surveys \& Tutorials}, vol.~21, no.~1, pp. 10--27, 2019.

\bibitem{Jianhua2024}
J.~Zhang, J.~Lin, P.~Tang, W.~Fan, Z.~Yuan, X.~Liu, H.~Xu, Y.~Lyu, L.~Tian, and P.~Zhang, ``{Deterministic Ray Tracing: A Promising Approach to {THz} Channel Modeling in {6G} Deployment Scenarios},'' \emph{IEEE Communications Magazine}, vol.~62, no.~2, pp. 48--54, 2024.

\bibitem{Yi2024}
H.~Yi, D.~He, P.~T. Mathiopoulos, B.~Ai, J.~M. García-Loygorri, J.~Dou, and Z.~Zhong, ``{Ray Tracing Meets Terahertz: Challenges and Opportunities},'' \emph{IEEE Communications Magazine}, vol.~62, no.~2, pp. 40--46, 2024.

\bibitem{Arnold2022}
M.~Arnold, M.~Bauhofer, S.~Mandelli, M.~Henninger, F.~Schaich, T.~Wild, and S.~ten Brink, ``{MaxRay: A Raytracing-based Integrated Sensing and Communication Framework},'' in \emph{2022 2nd IEEE International Symposium on Joint Communications \& Sensing (JC\&S)}, 2022, pp. 1--7.

\bibitem{Yiwen2021}
Y.~Wu, K.~Zhang, and Y.~Zhang, ``{Digital Twin Networks: A Survey},'' \emph{IEEE Internet of Things Journal}, vol.~8, no.~18, pp. 13\,789--13\,804, 2021.

\bibitem{Ruah2024}
C.~Ruah, O.~Simeone, J.~Hoydis, and B.~Al-Hashimi, ``{Calibrating Wireless Ray Tracing for Digital Twinning Using Local Phase Error Estimates},'' \emph{IEEE Transactions on Machine Learning in Communications and Networking}, vol.~2, pp. 1193--1215, 2024.

\bibitem{R12400691}
3GPP, ``{3rd Generation Partnership Project; Technical Specification Group Radio Access Network; Channel modelling for integrated sensing and communication with NR},'' 3rd Generation Partnership Project (3GPP), Technical Specification (TS) R1-2400691 (2024-03), 2024.

\bibitem{oran_digital_twin_2025}
\BIBentryALTinterwordspacing
{O-RAN Alliance}, ``{Research Report on Digital Twin RAN Use Cases},'' {O-RAN Alliance}, Tech. Rep., March 2025, accessed: 2025-03-06. [Online]. Available: \url{https://www.o-ran.org/research-reports/digital-twin-ran-use-cases}
\BIBentrySTDinterwordspacing

\bibitem{Klautau2018}
A.~Klautau, P.~Batista, N.~González-Prelcic, Y.~Wang, and R.~W. Heath, ``{5G MIMO Data for Machine Learning: Application to Beam-Selection Using Deep Learning},'' in \emph{2018 Information Theory and Applications Workshop (ITA)}, 2018, pp. 1--9.

\bibitem{Novak2021}
\BIBentryALTinterwordspacing
R.~Novak, A.~Hrovat, M.~D. Bedford, and T.~Javornik, ``{Geometric Simplifications of Natural Caves in Ray-Tracing-Based Propagation Modelling},'' \emph{Electronics}, vol.~10, no.~23, 2021. [Online]. Available: \url{https://www.mdpi.com/2079-9292/10/23/2914}
\BIBentrySTDinterwordspacing

\bibitem{Mozart2024}
\BIBentryALTinterwordspacing
L.~Mozart, J.~Borges, I.~Correa, and A.~Klautau, ``{Adjusting the Level of Detail in {3D} Models for Wireless Channel Generation Using Ray-tracing},'' in \emph{XLII Brazilian Symposium on Telecommunications and Signal Processing}.\hskip 1em plus 0.5em minus 0.4em\relax Sociedade Brasileira de Telecomunica{\c c}{\~o}es, 2024. [Online]. Available: \url{http://dx.doi.org/10.14209/sbrt.2024.1571036262}
\BIBentrySTDinterwordspacing

\bibitem{Zentner2013}
R.~Zentner and A.~K. Mucalo, ``{Ray Tracing Interpolation for Continuous Modeling of Double Directional Radio Channel},'' in \emph{Eurocon 2013}, 2013, pp. 212--217.

\bibitem{zhu2024}
M.~Zhu, L.~Cazzella, F.~Linsalata, M.~Magarini, M.~Matteucci, and U.~Spagnolini, ``{Toward Real-Time Digital Twins of EM Environments: Computational Benchmark for Ray Launching Software},'' \emph{IEEE Open Journal of the Communications Society}, vol.~5, pp. 6291--6302, 2024.

\bibitem{Hussain2017}
S.~Hussain and C.~Brennan, ``{An Efficient Ray Tracing Method for Propagation Prediction Along a Mobile Route in Urban Environments},'' \emph{Radio Science}, vol.~52, no.~7, pp. 862--873, 2017.

\bibitem{Borges2024}
J.~Borges, F.~Bastos, I.~Correa, P.~Batista, and A.~Klautau, ``{{CAVIAR}: Co-Simulation of {6G} Communications, {3-D} Scenarios, and {AI} for Digital Twins},'' \emph{IEEE Internet of Things Journal}, vol.~11, no.~19, pp. 31\,287--31\,300, 2024.

\bibitem{Testolina2024}
\BIBentryALTinterwordspacing
P.~Testolina, M.~Polese, P.~Johari, and T.~Melodia, ``{Boston Twin: the Boston Digital Twin for Ray-Tracing in {6G} Networks},'' in \emph{Proceedings of the 15th ACM Multimedia Systems Conference}, ser. MMSys '24.\hskip 1em plus 0.5em minus 0.4em\relax New York, NY, USA: Association for Computing Machinery, 2024, p. 441–447. [Online]. Available: \url{https://doi.org/10.1145/3625468.3652190}
\BIBentrySTDinterwordspacing

\bibitem{yu2025}
\BIBentryALTinterwordspacing
L.~Yu, Y.~Miao, J.~Zhang, S.~Liu, Y.~Zhang, and G.~Liu, ``{Road to 6G Digital Twin Networks: Multi-Task Adaptive Ray-Tracing as a Key Enabler},'' 2025. [Online]. Available: \url{https://arxiv.org/abs/2502.14290}
\BIBentrySTDinterwordspacing

\bibitem{woong2022}
\BIBentryALTinterwordspacing
W.~Seo, S.~Park, and I.~Ihm, ``{Efficient Ray Tracing of Large {3D} Scenes for Mobile Distributed Computing Environments},'' \emph{Sensors}, vol.~22, no.~2, 2022. [Online]. Available: \url{https://www.mdpi.com/1424-8220/22/2/491}
\BIBentrySTDinterwordspacing

\bibitem{parker2010}
\BIBentryALTinterwordspacing
S.~G. Parker, J.~Bigler, A.~Dietrich, H.~Friedrich, J.~Hoberock, D.~Luebke, D.~McAllister, M.~McGuire, K.~Morley, A.~Robison, and M.~Stich, ``{OptiX: a General Purpose Ray Tracing Engine},'' \emph{ACM Trans. Graph.}, vol.~29, no.~4, jul 2010. [Online]. Available: \url{https://doi.org/10.1145/1778765.1778803}
\BIBentrySTDinterwordspacing

\bibitem{Heath2016}
R.~W. Heath, N.~González-Prelcic, S.~Rangan, W.~Roh, and A.~M. Sayeed, ``{An Overview of Signal Processing Techniques for Millimeter Wave {MIMO} Systems},'' \emph{IEEE Journal of Selected Topics in Signal Processing}, vol.~10, no.~3, pp. 436--453, 2016.

\bibitem{Trindade2018}
\BIBentryALTinterwordspacing
I.~Trindade, B.~Boas, and A.~Klautau, ``{Evaluation of Simplified Methodology for Obtaining mmWave {MIMO} Channels from Ray-Tracing Simulations},'' in \emph{XXXVI Brazilian Symposium on Telecommunications and Signal Processing}.\hskip 1em plus 0.5em minus 0.4em\relax Sociedade Brasileira de Telecomunica{\c c}{\~o}es, 2018. [Online]. Available: \url{http://dx.doi.org/10.14209/sbrt.2018.341}
\BIBentrySTDinterwordspacing

\bibitem{Gotszald2015}
\BIBentryALTinterwordspacing
Łukasz Piotr~Gotszald, ``{Novel Tracing Algorithm vs {Remcom Wireless InSite}},'' \emph{International Journal of Electronics and Telecommunications}, vol.~61, pp. 273--279, 2015. [Online]. Available: \url{https://api.semanticscholar.org/CorpusID:56310618}
\BIBentrySTDinterwordspacing

\bibitem{wirelessInsite}
``{W}ireless {I}n{S}ite® 3{D} {W}ireless {P}rediction {S}oftware {O}verview | {R}emcom --- remcom.com,'' \url{https://www.remcom.com/resources/brochures/wireless-insite-overview}, 2024, [Accessed 28-11-2024].

\bibitem{Hoydis2023}
\BIBentryALTinterwordspacing
J.~Hoydis, S.~Cammerer, F.~A. Aoudia, A.~Vem, N.~Binder, G.~Marcus, and A.~Keller, ``{Sionna: An Open-Source Library for Next-Generation Physical Layer Research},'' 2023. [Online]. Available: \url{https://arxiv.org/abs/2203.11854}
\BIBentrySTDinterwordspacing

\bibitem{3gppTR38901}
\BIBentryALTinterwordspacing
{{3rd Generation Partnership Project (3GPP)}}, ``{{Study on Channel Model for Frequencies from 0.5 to 100 {GHz}}},'' {3GPP}, Technical Report TR 38.901, June 2022, version 17.0.0, Section 7.6, pp. 55--60. [Online]. Available: \url{https://www.etsi.org/deliver/etsi_tr/138900_138999/138901/16.01.00_60/tr_138901v160100p.pdf}
\BIBentrySTDinterwordspacing

\bibitem{unrealengine}
{Epic Games, Inc}, ``{{U}nreal {E}ngine},'' \url{https://www.unrealengine.com/en-US}, 2025, [Accessed 14-03-2025].

\bibitem{unityUnityRealTime}
{Unity Technologies}, ``{U}nity {R}eal-{T}ime {D}evelopment {P}latform | 3{D}, 2{D}, {V}{R} \& {A}{R} {E}ngine --- unity.com,'' \url{https://unity.com/}, 2025, [Accessed 14-03-2025].

\bibitem{sharma2022}
S.~Sharma and V.~Kumar, ``\BIBforeignlanguage{en}{{A Comprehensive Review on Multi-objective Optimization Techniques: Past, Present and Future}},'' \emph{\BIBforeignlanguage{en}{Arch. Comput. Methods Eng.}}, vol.~29, no.~7, pp. 5605--5633, Nov. 2022.

\bibitem{Bedford2020}
M.~D. Bedford, A.~Hrovat, G.~Kennedy, T.~Javornik, and P.~Foster, ``{Modeling Microwave Propagation in Natural Caves Using {LiDAR} and Ray Tracing},'' \emph{IEEE Transactions on Antennas and Propagation}, vol.~68, no.~5, pp. 3878--3888, 2020.

\bibitem{Mi2020}
H.~Mi, D.~He, K.~Guan, B.~Ai, C.~Liu, T.~Shui, L.~Zhu, and H.~Mei, ``{Implementation and Evaluation of Ray-Tracing Acceleration Methods in Wireless Communication},'' in \emph{2020 14th European Conference on Antennas and Propagation (EuCAP)}, 2020, pp. 1--5.

\bibitem{Novak2016}
R.~Novak, ``{Discrete Method of Images for {3D} Radio Propagation Modeling},'' \emph{3D Res.}, vol.~7, no.~3, Sep. 2016.

\bibitem{Liu2024}
\BIBentryALTinterwordspacing
Z.~Liu, P.~Zhao, L.~Guo, Z.~Nan, Z.~Zhong, and J.~Li, ``{Three-Dimensional Ray-Tracing-Based Propagation Prediction Model for Macrocellular Environment at Sub-6 {GHz} Frequencies},'' \emph{Electronics}, vol.~13, no.~8, 2024. [Online]. Available: \url{https://www.mdpi.com/2079-9292/13/8/1451}
\BIBentrySTDinterwordspacing

\bibitem{Lecci2020}
M.~Lecci, P.~Testolina, M.~Giordani, M.~Polese, T.~Ropitault, C.~Gentile, N.~Varshney, A.~Bodi, and M.~Zorzi, ``{Simplified Ray Tracing for the Millimeter Wave Channel: A Performance Evaluation},'' in \emph{2020 Information Theory and Applications Workshop (ITA)}, 2020, pp. 1--6.

\bibitem{Lecci2021}
M.~Lecci, P.~Testolina, M.~Polese, M.~Giordani, and M.~Zorzi, ``{Accuracy Versus Complexity for mmWave Ray-Tracing: A Full Stack Perspective},'' \emph{IEEE Transactions on Wireless Communications}, vol.~20, no.~12, pp. 7826--7841, 2021.

\bibitem{qdrealization}
M.~Lecci and P.~Testolina, ``{Q}uasi-{D}eterministic {C}hannel {R}ealization {S}oftware --- nist.gov,'' \url{https://www.nist.gov/services-resources/software/quasi-deterministic-channel-realization-software}, 2021, [Accessed 28-11-2024].

\bibitem{blender}
\BIBentryALTinterwordspacing
{Blender Online Community}, ``{Blender - a {3D} Modelling and Rendering Package},'' Blender Foundation, Stichting Blender Foundation, Amsterdam, 2018. [Online]. Available: \url{http://www.blender.org}
\BIBentrySTDinterwordspacing

\bibitem{Low1997}
\BIBentryALTinterwordspacing
K.-L. Low and T.-S. Tan, ``{Model Simplification Using Vertex-clustering},'' in \emph{Proceedings of the 1997 Symposium on Interactive 3D Graphics}, ser. I3D '97.\hskip 1em plus 0.5em minus 0.4em\relax New York, NY, USA: Association for Computing Machinery, 1997, p. 75–ff. [Online]. Available: \url{https://doi.org/10.1145/253284.253310}
\BIBentrySTDinterwordspacing

\bibitem{Berg2008}
M.~de~Berg, O.~Cheong, M.~van Kreveld, and M.~Overmars, \emph{Computational geometry}, 3rd~ed.\hskip 1em plus 0.5em minus 0.4em\relax Springer Berlin Heidelberg, Mar. 2008.

\bibitem{Cignoni1998}
\BIBentryALTinterwordspacing
P.~Cignoni, C.~Montani, and R.~Scopigno, ``{A Comparison of Mesh Simplification Algorithms},'' \emph{Computers \& Graphics}, vol.~22, no.~1, pp. 37--54, 1998. [Online]. Available: \url{https://www.sciencedirect.com/science/article/pii/S0097849397000824}
\BIBentrySTDinterwordspacing

\bibitem{Li2021}
Y.~Li, N.~Li, and C.~Han, ``{Ray-tracing Simulation and Hybrid Channel Modeling for Low-Terahertz UAV Communications},'' in \emph{ICC 2021 - IEEE International Conference on Communications}, 2021, pp. 1--6.

\bibitem{cignoni2008meshlab}
P.~Cignoni, M.~Callieri, M.~Corsini, M.~Dellepiane, F.~Ganovelli, G.~Ranzuglia \emph{et~al.}, ``{Meshlab: an Open-source Mesh Processing Tool.}'' in \emph{Eurographics Italian chapter conference}, vol. 2008.\hskip 1em plus 0.5em minus 0.4em\relax Salerno, Italy, 2008, pp. 129--136.

\bibitem{garland1997}
\BIBentryALTinterwordspacing
M.~Garland and P.~S. Heckbert, ``{Surface Simplification Using Quadric Error Metrics},'' in \emph{Proceedings of the 24th Annual Conference on Computer Graphics and Interactive Techniques}, ser. SIGGRAPH '97.\hskip 1em plus 0.5em minus 0.4em\relax USA: ACM Press/Addison-Wesley Publishing Co., 1997, p. 209–216. [Online]. Available: \url{https://doi.org/10.1145/258734.258849}
\BIBentrySTDinterwordspacing

\bibitem{rossignac1993multi}
J.~Rossignac and P.~Borrel, ``{Multi-resolution {3D} Approximations for Rendering Complex Scenes},'' in \emph{Modeling in computer graphics: methods and applications}.\hskip 1em plus 0.5em minus 0.4em\relax Springer Berlin Heidelberg, 1993, pp. 455--465.

\bibitem{rossignac1997geometric}
J.~Rossignac, ``{Geometric Simplification and Compression in Multiresolution Surface Modeling},'' in \emph{SIGGRAPH Course Notes \#25}, 1997.

\bibitem{Wu2022}
\BIBentryALTinterwordspacing
K.~Wu, X.~He, Z.~Pan, and X.~Gao, ``{Occluder Generation for Buildings in Digital Games},'' \emph{Computer Graphics Forum}, vol.~41, no.~7, pp. 205--214, 2022. [Online]. Available: \url{https://onlinelibrary.wiley.com/doi/abs/10.1111/cgf.14669}
\BIBentrySTDinterwordspacing

\bibitem{sionna_rt_docs}
{NVIDIA corporation}, ``{R}ay {T}racing - {S}ionna 0.19.2 documentation --- nvlabs.github.io,'' \url{https://jhoydis.github.io/sionna-0.19.2-doc/}, 2025, [Accessed 09-04-2025].

\end{thebibliography}
\newpage
\begin{IEEEbiography}
[{\includegraphics[width=1in,height=1.15in,clip]{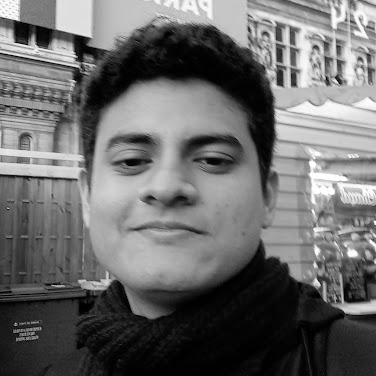}}]{Cláudio Modesto} received a B.Sc. degree in computer engineering from the Federal University of Pará (UFPA), Belém, Pará, Brazil, in 2024. He is currently pursuing an M.Sc. degree in electrical engineering at UFPA. He has been part of the Research and Development Center for Telecommunications, Automation, and Electronics (LASSE) since 2023. His current research interests include signal processing for wireless communication, machine learning, and deep learning applied to graphs.
\end{IEEEbiography}
\begin{IEEEbiography}[{\includegraphics[width=1in,height=1.1in]{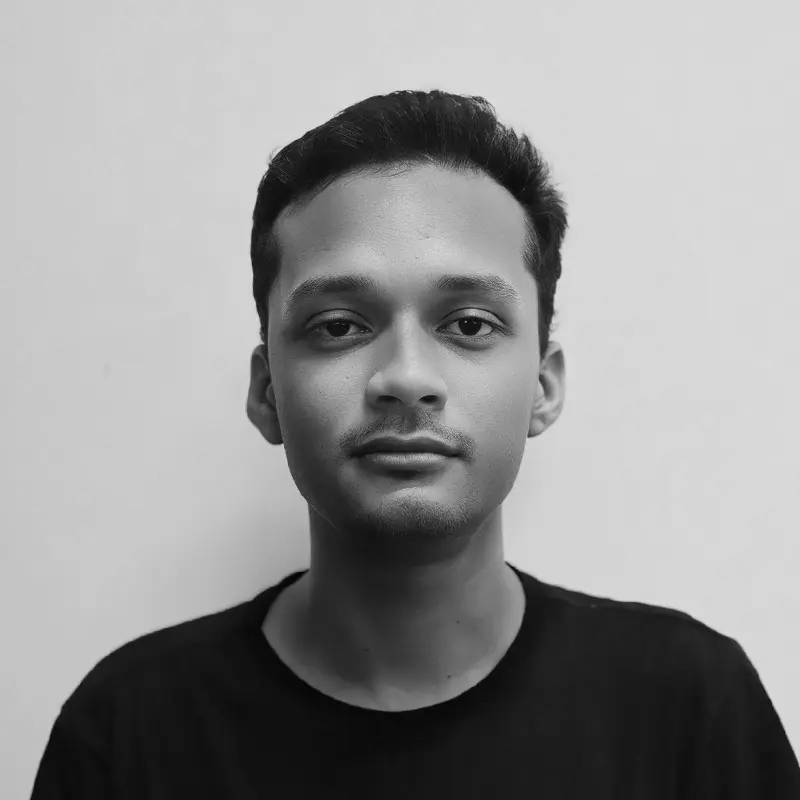}}]{Lucas Mozart} is currently pursuing a B.Sc. degree in computer engineering at the Federal University of Pará (UFPA). He has been a member of the Research and Development Center for Telecommunications, Automation, and Electronics (LASSE) since 2023. His current research interests are machine learning applied to telecommunications.
\end{IEEEbiography}
\begin{IEEEbiography}[{\includegraphics[width=1in,height=1.25in,clip,keepaspectratio]{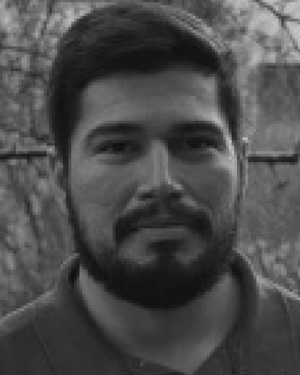}}]{Pedro Batista} received his B.Sc., M.Sc. and Ph.D. degrees from the Electrical Engineering Graduate Program of the Federal University of Pará, Brazil. He is currently a researcher with Ericsson. His research interests are in the optimization of future mobile networks, particularly using machine learning and machine reasoning, and future Internet architectures.
\end{IEEEbiography}
\begin{IEEEbiography}[{\includegraphics[width=1in,height=1.25in,clip,keepaspectratio]{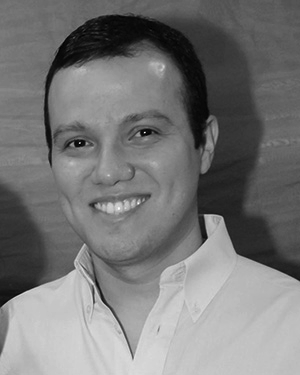}}]{André Cavalcante} (Member, IEEE) received a Ph.D. degree in electrical engineering from the Federal University of Pará (UFPA), Brazil, in 2007. From 2007 to 2016, he was a co-worker at the Nokia Institute of Technology (INDT) in Brazil, where he was involved in several research and development projects related to wireless communications. He is currently a Senior Researcher with Ericsson Research in Brazil. His research interests include mobile transport networks for future wireless communication systems.
\end{IEEEbiography}
\begin{IEEEbiography}[{\includegraphics[width=1in,height=1.25in,clip,keepaspectratio]{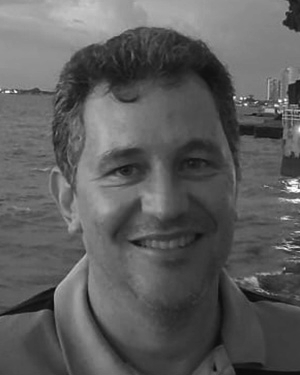}}]{Aldebaro Klautau} (Senior Member, IEEE)
received a bachelor's degree (Federal University of Pará, UFPA, 1990), M.Sc. (Federal Univ. of Santa Catarina, UFSC, 1993) and Ph.D. degrees (University of California at San Diego, UCSD, 2003) in electrical engineering. He is a full professor at UFPA, where he is the ITU Focal Point and coordinates the LASSE Research Group. He is a researcher at CNPq, Brazil, and is a senior member of the IEEE and the Brazilian Telecommunications Society (SBrT). His work focuses on machine learning and signal processing for communication.
\end{IEEEbiography}







\end{document}